\documentclass[10pt,twocolumn,brazil,english,manuscript]{revtex4}
\usepackage[T1]{fontenc}
\usepackage[latin9]{inputenc}
\usepackage{amsmath}
\usepackage{amssymb}
\usepackage{graphicx}

\makeatletter
\@ifundefined{textcolor}{}
{%
 \definecolor{BLACK}{gray}{0}
 \definecolor{WHITE}{gray}{1}
 \definecolor{RED}{rgb}{1,0,0}
 \definecolor{GREEN}{rgb}{0,1,0}
 \definecolor{BLUE}{rgb}{0,0,1}
 \definecolor{CYAN}{cmyk}{1,0,0,0}
 \definecolor{MAGENTA}{cmyk}{0,1,0,0}
 \definecolor{YELLOW}{cmyk}{0,0,1,0}
 }

\makeatother

\usepackage{babel}
\begin{document}

\title{Thermal entanglement in exactly solvable Ising-XXZ diamond chain structure}

\author{Onofre Rojas$^{1}$, M Rojas$^{1}$, N. S. Ananikian$^{2}$ and S.
M. de Souza$^{1}$}

\affiliation{$^{1}$Departamento de Ciencias Exatas, Universidade Federal de Lavras,
37200-000, Lavras-MG, Brazil}

\affiliation{$^{2}$A.I. Alikhanyan National Science Laboratory, 0036 Yerevan,
Armenia.}
\begin{abstract}
Most quantum entanglement investigations are focused on two qubits or some finite (small) chain structure, since the infinite chain structure is a considerably  cumbersome task. Therefore, the quantum entanglement properties involving an infinite chain structure is quite important, not only because the mathematical calculation is cumbersome but also because real materials are well represented by an infinite chain. Thus, in this paper we consider an entangled diamond chain with Ising and anisotropic Heisenberg (Ising-XXZ) coupling. Two
interstitial particles are coupled through Heisenberg coupling or
simply two-qubit Heisenberg, which could be responsible for the
emergence of entanglement. These two-qubit Heisenberg operators are interacted with
two nodal Ising spins. An infinite diamond chain is organized by
interstitial- interstitial and nodal-interstitial (dimer-monomer)
site couplings. We are able to get the thermal average of the two-qubit
operator, called the reduced two-qubit density operator. Since these
density operators are spatially separated, we could obtain the
concurrence (entanglement) directly in the thermodynamic limit. The
thermal entanglement (concurrence) is constructed for different
values of the anisotropic Heisenberg parameter, magnetic field and
temperature. We also observed the threshold temperature via the
parameter of anisotropy, Heisenberg and Ising interaction, external
magnetic field, and temperature.
\end{abstract}
\maketitle

\section{Introduction}

Quantum entanglement is one of the most attractive types of
correlations that can be shared only among quantum
systems \cite{AmicoHorod}. In recent years, many efforts have been
devoted to characterize qualitatively and quantitatively the
entanglement properties of condensed matter systems, which are the
natural candidate to apply for quantum communication and quantum
information. In this sense, it is very important to study the
entanglement of solid state systems such as spin
chains \cite{qubit-Heisnb}. The Heisenberg chain is one of the
simplest quantum systems, which exhibits entanglement; due to
the Heisenberg interaction it is not localized in the spin system.

 In the last decade, several diamond chain structures have been
discussed. It is interesting to consider the quantum
antiferromagnetic Heisenberg model on a generalized  diamond chain,
which describes real materials such as
$\mathrm{Cu_{3}(CO_{3})_{2}(OH)_{2}}$, known as natural azurite.
Honecker et al. \cite{honecker} studied the dynamic and
thermodynamic properties for this model. In addition, Pereira et al.
\cite{Pereira08,pereira09} investigated the magnetization plateau of
delocalized interstitial spins, as well as magnetocaloric effect in
kinetically frustrated diamond chains. More recently, Lisnii
\cite{lisnii} studied a distorted diamond Ising-Hubbard chain, and
that the model also reveals the geometrical frustration.
Thermodynamics of the Ising-Heisenberg model on a diamond-like chain
was widely discussed in the Refs
\cite{canova06,vadim,valverde,lisnii-11}.

The motivation for the study of the Ising and anisotropic Heisenberg
(Ising-XXZ) model on a diamond chain, according to  experiments of the
natural mineral azurite, theoretical calculations of the Ising-XXZ
model, as well as the experimental result of the dimer (interstitial
sites) exchange parameter is based on a number of recent works. The
1/3 magnetization plateau and the double peaks both in the magnetic
susceptibility and specific heat were observed in the experimental
measurements \cite{rule,kikuchi}. It should be noticed that the dimer
(interstitial sites)  exchange is much  stronger than those
nodal sites.  Various types of theoretical Heisenberg-model
approximate methods were proposed:  renormalization of the
density matrix renormalization group of the transfer matrix, 
density functional theory,  high-temperature expansion, 
variational mean-field-like treatment based on the Gibbs-Bogoliubov
inequality, and  Lanczos diagonalization on a diamond chain to explain
the experimental measurements (magnetization plateau and the double
peaks) in the natural mineral azurite \cite{honecker-lauchli}. The
localizable entanglement (LE) has been calculated for ground states
of arbitrary Hamiltonians numerically by making use of the density
renormalization group formalism and exhibited characteristic
features at a quantum phase transition. The LE is completely
characterized by the maximal connected correlation function for
ground states of spin-1/2 systems \cite{martin}. All of these
theoretical studies are approximate. There is another possibility.
Since dimer interaction is much higher than the rest, it can be
represented as an exactly solvable Ising-Heisenberg model. In
addition, experimental data on the magnetization plateau coincide
with the approximation Ising-Heisenberg 
model \cite{canova06,Pereira08,ananikian,chakh}.

Several studies have been done on the threshold temperature for the
pairwise thermal entanglement in the Heisenberg model with a finite
number of qubits. Thermal entanglement of the isotropic Heisenberg
chain of spin has been studied in the absence \cite{wang} and in the
presence of an external magnetic field \cite{X-Wang,arnesen}. The
pairwise thermal entanglement of nearest-neighbor qubits is
independent of the sign of exchange constants and the sign of
magnetic fields in the XX even-number qubit ring with a magnetic
field. The thermal entanglement in
Heisenberg XX decreases with increasing temperature and the
threshold temperature value $T_{th}$ is independent of the external
magnetic field. Although in some references the threshold temperature
$T_{th}$ is called the quantum critical temperature, here we call it
the threshold temperature to avoid misunderstanding with
the magnetic phase transition temperature.

It is quite relevant to study the thermal entanglement of the
Ising-XXZ on diamond chain, as Ananikian et al.
\cite{ananikian,chakh} have discussed  the thermal entanglement of
the Ising-Heisenberg chain in the isotropic limit in the cluster
approach. The paper is organized as follows: In Sec. II we present
the Ising-XXZ model on diamond chain. We obtain the exact
solution of the model via the transfer-matrix approach and its dimer
(two-qubit) reduced density operator in Sec. III. In Sec. IV, we
discuss the thermal entanglement of the Heisenberg reduced density
operator of the model, such as concurrence and threshold
temperature. Finally, concluding remarks are given in Sec. V.

\section{Ising-XXZ diamond Hamiltonian}

\begin{figure}
\includegraphics[scale=0.52]{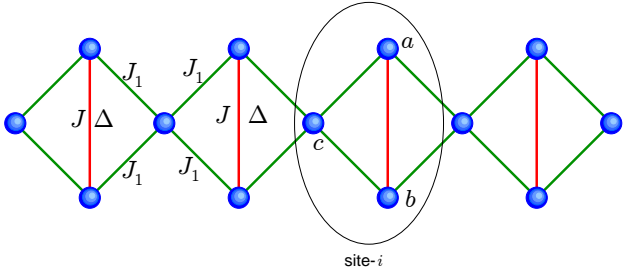}\caption{\label{fig:diamond}(Color online) Schematic representation of XXZ-Ising diamond
chain. Red line represents the quantum bipartite coupling.}
\end{figure}

In this work, we study the Ising-XXZ model with mixed nodal Ising spins
and interstitial anisotropic Heisenberg spins in the presence of an
external magnetic field on a diamond-like chain.   The thermodynamic
properties were previously discussed in Refs.
\cite{canova06,valverde,lisnii,vadim}. A diamond-like chain is
schematically illustrated  in Fig 1. The Hamiltonian operator can
be expressed as follows:
\begin{align}
\mathcal{H}= & \sum_{i=1}^{N}J\left(\boldsymbol{S}_{a,i},\boldsymbol{S}_{b,i}\right)_{\Delta}+J_{1}\left(S_{a,i}^{z}+S_{b,i}^{z}\right)\left(\mu_{i}+\mu_{i+1}\right)+\nonumber \\
 & -h_{0}\left(S_{a,i}^{z}+S_{b,i}^{z}\right)-\frac{h}{2}\left(\mu_{i}+\mu_{i+1}\right),\label{eq:Hamil-1}
\end{align}
where
$\left(\boldsymbol{S}_{a,i},\boldsymbol{S}_{b,i}\right)_{\Delta}=S_{a,i}^{x}S_{b,i}^{x}+S_{a,i}^{y}S_{b,i}^{y}+\Delta
S_{a,i}^{z}S_{b,i}^{z}$ corresponds to  the interstitial
anisotropic Heisenberg spins coupling ($J$ and $\Delta$), while the
nodal-interstitial (dimer-monomer) spins are representing by Ising-type
 exchanges ($J_{1}$). The Hamiltonian also includes a longitudinal external
 magnetic field $h_{0}$ acting on Heisenberg spins and a magnetic  $h$ acting on Ising spins.
For convenience, we will consider the case $h_{0}=h$.

The quantum Heisenberg spin coupling can be expressed using
matrix notation.  We have
\begin{equation}
\left(\boldsymbol{S}_{a,i},\boldsymbol{S}_{b,i}\right)_{\Delta}=\left[\begin{array}{cccc}
\frac{\Delta}{4} & 0 & 0 & 0\\
0 & -\frac{\Delta}{4} & \frac{1}{2} & 0\\
0 & \frac{1}{2} & -\frac{\Delta}{4} & 0\\
0 & 0 & 0 & \frac{\Delta}{4}
\end{array}\right]
\end{equation}
and
\begin{equation}
S_{a,i}^{z}+S_{b,i}^{z}=\left[\begin{array}{cccc}
1 & 0 & 0 & 0\\
0 & 0 & 0 & 0\\
0 & 0 & 0 & 0\\
0 & 0 & 0 & -1
\end{array}\right].
\end{equation}

We obtain the following eigenvalues by diagonalization of the
two-qubit Heisenberg (dimer) exchange and by assuming fixed values for $\mu_{i}$ and $\mu_{i+1}$:
\begin{align}
\mathcal{E}_{1}(\mu_{i},\mu_{i+1})= & \frac{J\Delta}{4}+\left(J_{1}-\frac{h}{2}\right)\left(\mu_{i}+\mu_{i+1}\right)-h,\nonumber \\
\mathcal{E}_{2}(\mu_{i},\mu_{i+1})= & \frac{J}{2}-\frac{J\Delta}{4}-\frac{h}{2}\left(\mu_{i}+\mu_{i+1}\right),\nonumber \\
\mathcal{E}_{3}(\mu_{i},\mu_{i+1})= & -\frac{J}{2}-\frac{J\Delta}{4}-\frac{h}{2}\left(\mu_{i}+\mu_{i+1}\right),\nonumber \\
\mathcal{E}_{4}(\mu_{i},\mu_{i+1})= & \frac{J\Delta}{4}-\left(J_{1}+\frac{h}{2}\right)\left(\mu_{i}+\mu_{i+1}\right)+h,\label{eq:eigenvals}
\end{align}
where their corresponding eigenstate in terms of standard basis $\{|\uparrow\uparrow\rangle,|\downarrow\downarrow\rangle,|\uparrow\downarrow\rangle,|\downarrow\uparrow\rangle\}$
are given respectively by
\begin{align}
|\varphi_{1}\rangle= & |\uparrow\uparrow\rangle,\\
|\varphi_{2}\rangle= & \frac{1}{\sqrt{2}}\left(|\uparrow\downarrow\rangle+|\downarrow\uparrow\rangle\right),\label{eq:phi2}\\
|\varphi_{3}\rangle= & \frac{1}{\sqrt{2}}\left(|\uparrow\downarrow\rangle-|\downarrow\uparrow\rangle\right),\label{eq:phi3}\\
|\varphi_{4}\rangle= & |\downarrow\downarrow\rangle.
\end{align}

In quantum information the states $|\varphi_{2}\rangle$ and
$|\varphi_{3}\rangle$ are known as the \textquotedbl{}magic
basis\textquotedbl{} or Bell state, where two-qubit Heisenberg are
maximally entangled. These states are responsible for the rise  of
entanglement at finite temperature.

In order to study the two-qubit entanglement, we use the
concurrence \cite{wootters,hill}, which is defined by
\begin{equation}
\mathcal{C}=|\langle\varphi_{i}|\tilde{\varphi}_{i}\rangle|,
\end{equation}
where $|\tilde{\varphi}_{i}\rangle$ means the spin-flip state.
Therefore, the concurrence $\mathcal{C}$ will be for states
$|\varphi_{2}\rangle$ and $|\varphi_{3}\rangle$, which corresponds
to qubits that are maximally entangled.  If the states
$|\varphi_{1}\rangle$ and $|\varphi_{4}\rangle$ have no concurrence
($\mathcal{C}=0$), that means the qubits are unentangled. Later we
give the definition of entanglement in more detail.

\subsection{Zero temperature phase diagram of entangled state}

In this section we study the phase diagram of the entangled states,
similar to the magnetic phase previously discussed in Refs. \cite{canova06,valverde}, where it was observed three magnetic states,
a frustrated (FRU) state, a ferrimagnetic (FIM) state, and a
ferromagnetic (FM) state. Although we only have two states
(entangled and unentangled),  we discuss here the entangled region
which is  closely related to the magnetic states. It can be expressed
the following states: \foreignlanguage{brazil}{
\begin{align}
\mid ENT\rangle & =\overset{N}{\underset{i=1}{\prod}}\mid\varphi_{3}\rangle_{i}\otimes\mid\mu\rangle_{i},\\
\mid UFI\rangle & =\overset{N}{\underset{i=1}{\prod}}\mid\varphi_{1}\rangle_{i}\otimes\mid-\rangle_{i},\\
\mid UFM\rangle & =\overset{N}{\underset{i=1}{\prod}}\mid\varphi_{1}\rangle_{i}\otimes\mid+\rangle_{i},
\end{align}
}where $\mid\mu\rangle_{i}$ stands for arbitrary values
($\mu=\pm1/2$) of the nodal spin in the $i$-th block.

According to the Bell states given in eqs.(\ref{eq:phi2}) and
(\ref{eq:phi3}), at zero temperature the entangled (ENT) state is
fully spanned in the frustrated state for zero magnetic field, while the
ferrimagnetic and ferromagnetic state is spanned by unentangled
states denoted by UFI and UFM respectively.

\begin{figure}
\includegraphics[scale=0.15]{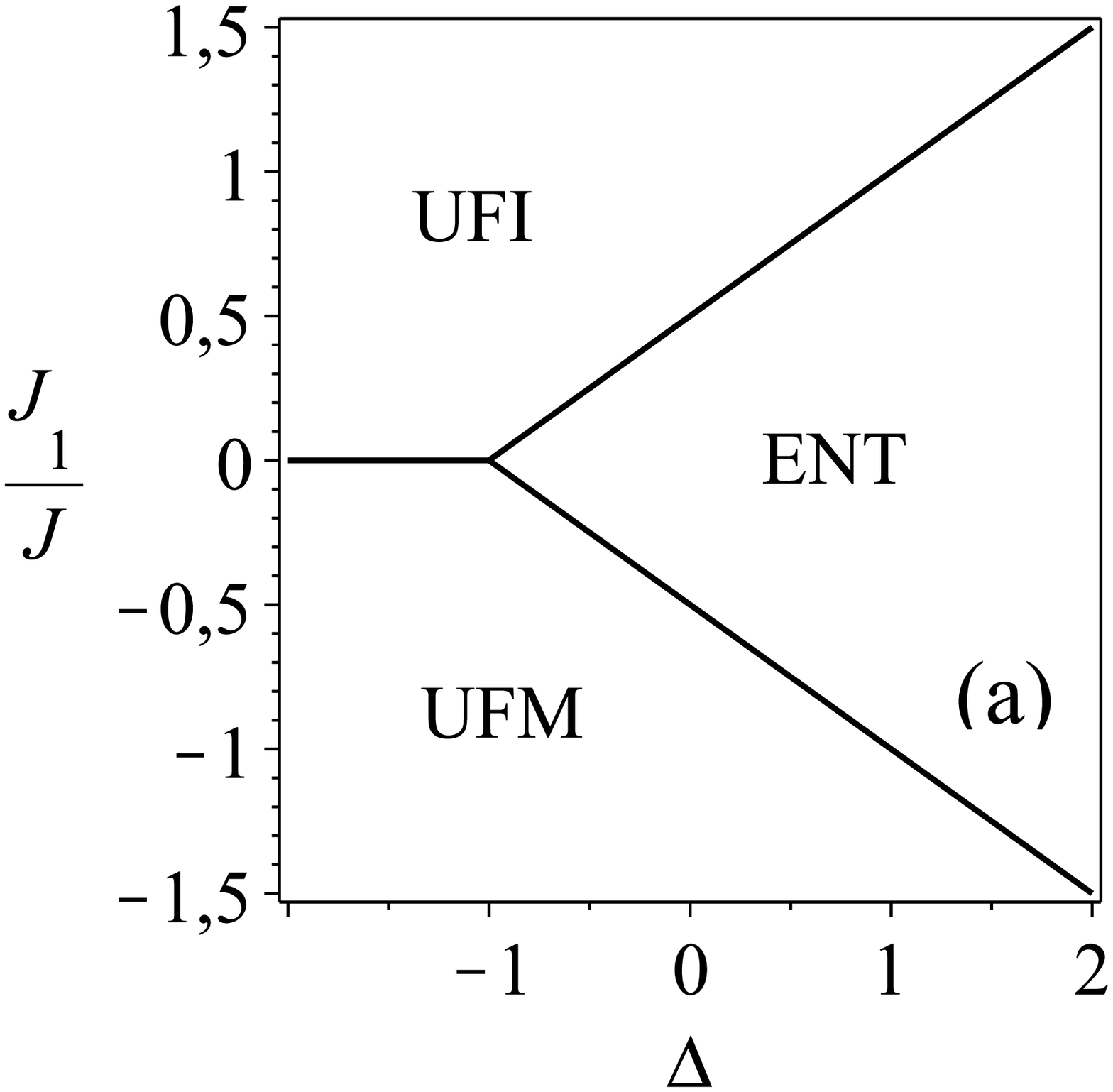}\includegraphics[scale=0.15]{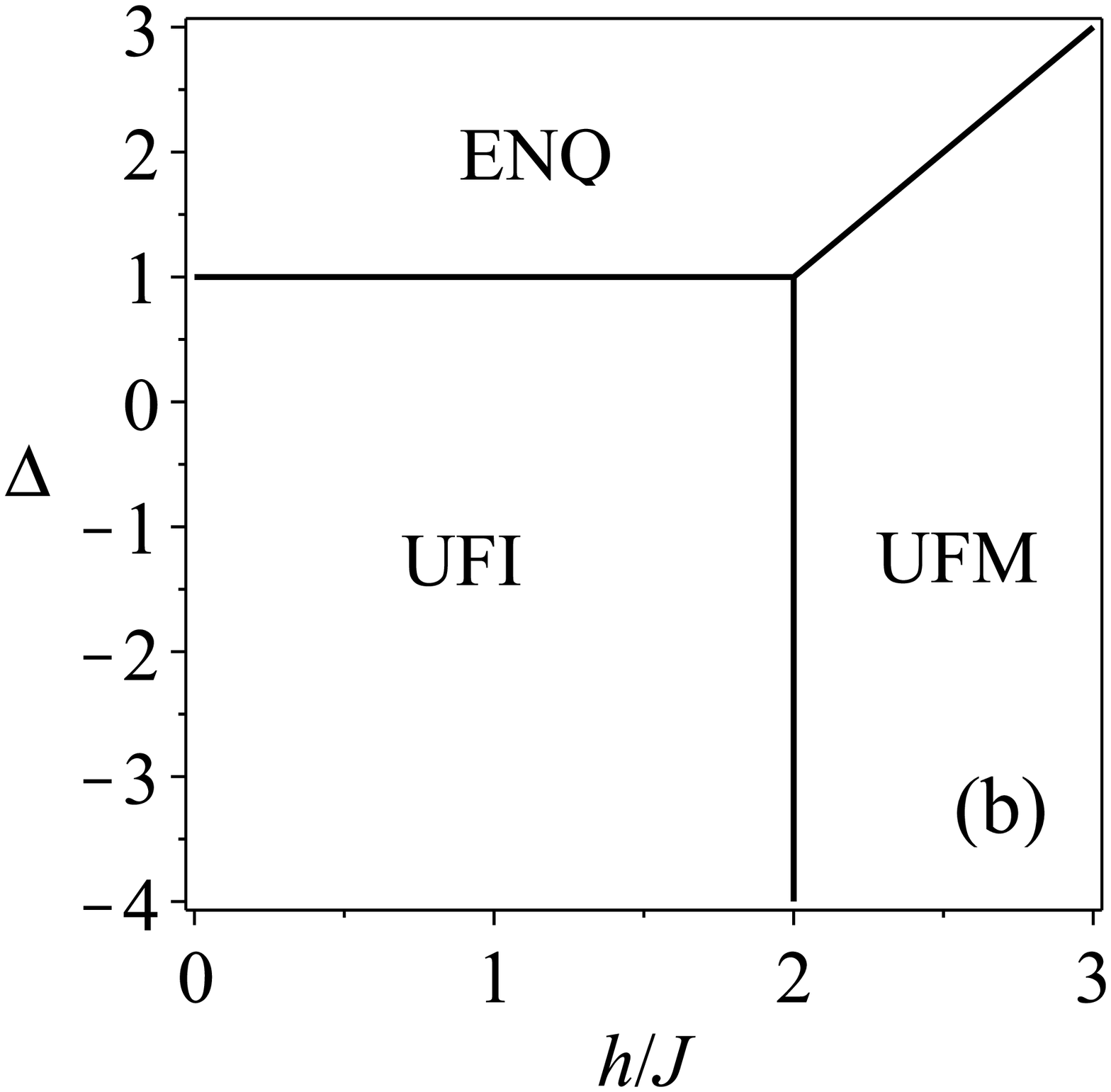}\includegraphics[scale=0.15]{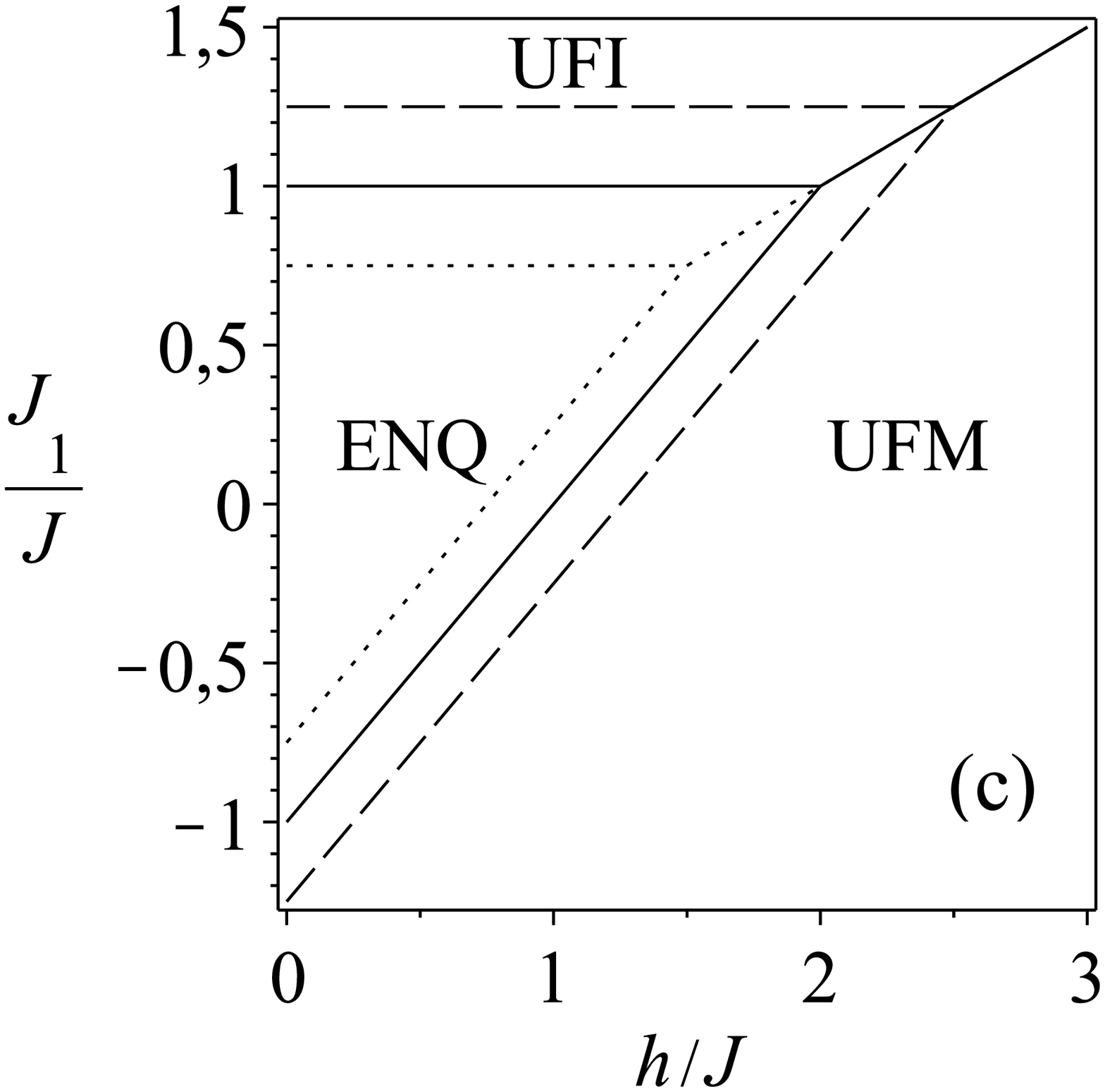}\caption{\label{fig:PHD}Phase diagrams of entangled state at zero temperature.
The entangled state in frustrated phase denoted by ENT, whereas the
entangled state in quantum ferrimagnetic phase we denote by ENQ. The
unentangled state in ferrimagnetic phase (UFI) and unentangled ferromagnetic
(UFM) states.}
\end{figure}

In Fig. \ref{fig:PHD}(a) we plot the phase diagram $\Delta$ versus $J_{1}/J$ at zero
temperature  for $h/J=h_{0}/J=0$ , where
the ground-state energy of the entangled state is
$E_{ENT}=-\frac{1}{2}-\frac{\Delta}{4}$, while for the unentangled
state (ferrimagnetic state) it is
$E_{UFI}=-\frac{J_{1}}{J}+\frac{\Delta}{4}$, and for the unentangled
state (ferromagnetic state) it is given by
$E_{UFM}=\frac{J_{1}}{J}+\frac{\Delta}{4}$. The phase boundary
between UFI and UFM is simply $J_{1}/J=0$, and the phase boundary
between the UFI and ENT states is given by
$(\frac{J_{1}}{J}=\frac{\Delta}{2}+\frac{1}{2})$, while the phase
boundary between UFM and ENT (frustrated) state is given by
$(\frac{J_{1}}{J}=-\frac{\Delta}{2}-\frac{1}{2})$.

In Fig. \ref{fig:PHD}(b) we plot the phase diagram $\Delta$ versus $h/J$ at zero temperature
for $J_{1}/J=1 $ and  $h/J=h_{0}/J$, where the eigenvalues for the unentangled
state UFM is given by $E_{UFM}=1+\frac{\Delta}{4}-\frac{3h}{2J}$,
while for the entangled state (quantum ferrimagnetic state) denoted by
$\mid ENQ\rangle=\overset{N}{\underset{i=1}{\prod}}\mid\varphi_{3}\rangle_{i}\otimes\mid+\rangle_{i},$
and its eigenvalues is $E_{ENQ}=-\frac{1}{2}-\frac{\Delta}{4}-\frac{h}{2J}$.
We also have the unentangled state UFI whose eigenvalue becomes $E_{UFI}=-1+\frac{\Delta}{4}-\frac{h}{2J}$.
The phase boundary between UFM and UFI is given by $h/J=2$, whereas
between the ENQ and UFI states it is limited by $\Delta=1$, while for the ENQ
and UFM states we have $\Delta=\frac{2h}{J}-3$.

Finally, in Fig. \ref{fig:PHD}(c)  we illustrate the phase diagram $J_{1}/J$ versus $h/J$
at zero temperature for $\Delta=1$ and $h/J=h_{0}/J$, where the eigenvalues
are given by $E_{ENQ}=-\frac{3}{4}-\frac{h}{2J}$ , $E_{UFI}=\frac{1}{4}-\frac{J_{1}}{J}-\frac{h}{2J}$
and $E_{UFM}=\frac{1}{4}+\frac{J_{1}}{J}-\frac{3h}{2J}$. The phase
boundary between the ENQ and UFI states is limited by $J_{1}/J=1$, while
the boundary between UFI and UFM is given by $J_{1}=h/2$, and between the
ENQ and UFM states the boundary follows the curve $\frac{J_{1}}{J}=\frac{h}{J}-1$.
The doted line corresponds to the phase diagram for $\Delta=0.5$, whereas
the dashed line represent the phase diagram for $\Delta=1.5$.

However, entangled states will change with increasing temperature.
This will be discussed in Sec. IV.

\section{The partition function and density operator}

With the aim of studying  any thermal quantities, we first need to
obtain a partition function on a diamond chain. As mentioned
earlier \cite{Fisher,Syozi,phys-A-09}, this model can be solved
exactly through a decoration transformation and transfer-matrix
approach \cite{baxter-book}. In order to summarize this approach we
will define the following operator as a function of Ising spin
particles $\mu$ and $\mu'$,

\begin{equation}
\varrho(\mu,\mu')=\mathrm{e}^{-\beta H_{ab}(\mu,\mu')},\label{eq:rho-loc}
\end{equation}
where $H_{ab}(\mu,\mu')$ corresponds to the $r$th-block Hamiltonian
(\ref{eq:Hamil-1}) (without summations), which depends on the
neighboring Ising spins $\mu$ and $\mu'$. Alternatively the operator
(\ref{eq:rho-loc}) could be written in terms of two-qubit operator
eigenvalues (\ref{eq:eigenvals}), which is
\begin{equation}
\varrho(\mu,\mu')=\sum_{i=1}^{4}\mathrm{e}^{-\beta\varepsilon_{i}(\mu,\mu')}|\varphi_{i}\rangle\langle\varphi_{i}|.
\end{equation}
Straightforwardly, we can obtain the Boltzmann factor by tracing out
over the two-qubit operator,
\begin{equation}
w(\mu,\mu')=\mathrm{tr}_{ab}\left(\varrho(\mu,\mu')\right)=\sum_{i=1}^{4}\mathrm{e}^{-\beta\varepsilon_{i}(\mu,\mu')},\label{eq:w-def}
\end{equation}
where the Ising-XXZ diamond chain partition function can be written
in terms of Boltzmann factors,
\begin{equation}
Z_{N}=\sum_{\{\mu\}}w(\mu_{1},\mu_{2})\ldots w(\mu_{N},\mu_{1}).
\end{equation}

Using the transfer-matrix notation, we can write the partition function
of the diamond chain straightforwardly by $Z_{N}=\mathrm{tr}\left(W^{N}\right),$
where the transfer-matrix is expressed as
\begin{equation}
W=\left[\begin{array}{cc}
w(\frac{1}{2},\frac{1}{2}) & w(\frac{1}{2},-\frac{1}{2})\\
w(-\frac{1}{2},\frac{1}{2}) & w(-\frac{1}{2},-\frac{1}{2})
\end{array}\right],\label{eq:W}
\end{equation}
where the transfer matrix elements are denoted by $w_{++}\equiv
w(\frac{1}{2},\frac{1}{2})$, $w_{+-}\equiv
w(\frac{1}{2},-\frac{1}{2})$ and $w_{--}\equiv
w(-\frac{1}{2},-\frac{1}{2})$.

After performing the diagonalization of the transfer matrix (\ref{eq:W}), the
eigenvalues are
\begin{equation}
\Lambda_{\pm}=\frac{w_{++}+w_{--}\pm Q}{2},
\end{equation}
assuming that $Q=\sqrt{\left(w_{++}-w_{--}\right)^{2}+4w_{+-}^{2}}$.
Therefore, the partition function for finite chain under periodic
boundary conditions is given by

\begin{equation}
Z_{N}=\Lambda_{+}^{N}+\Lambda_{-}^{N}.
\end{equation}

In the thermodynamic limit the partition function will be
simplified, which results in $Z_{N}=\Lambda_{+}^{N}$. It is worth
 noticing that this result was widely used in Refs.
\cite{canova06,valverde,vadim,lisnii}.

\subsection{Two-qubit operator }

In order to calculate the two-qubit Heisenberg operator bonded by
Ising particles $\mu$ and $\mu'$, we assume that the Ising
spin of the particle is fixed. Thus, the qubits operator elements
in the natural basis becomes
\begin{equation}
\varrho=\left[\begin{array}{cccc}
\varrho_{1,1} & 0 & 0 & 0\\
0 & \varrho_{2,2} & \varrho_{2,3} & 0\\
0 & \varrho_{3,2} & \varrho_{3,3} & 0\\
0 & 0 & 0 & \varrho_{4,4}
\end{array}\right],
\end{equation}

where the elements of the two-qubit operator are given by
\begin{align}
\varrho_{1,1}(\mu,\mu')= & \mathrm{e}^{-\beta\varepsilon_{1}(\mu,\mu')},\nonumber \\
\varrho_{2,2}(\mu,\mu')= & \frac{1}{2}\left(\mathrm{e}^{-\beta\varepsilon_{2}(\mu,\mu')}+\mathrm{e}^{-\beta\varepsilon_{3}(\mu,\mu')}\right),\nonumber \\
\varrho_{2,3}(\mu,\mu')= & \frac{1}{2}\left(\mathrm{e}^{-\beta\varepsilon_{2}(\mu,\mu')}-\mathrm{e}^{-\beta\varepsilon_{3}(\mu,\mu')}\right),\nonumber \\
\varrho_{4,4}(\mu,\mu')= & \mathrm{e}^{-\beta\varepsilon_{4}(\mu,\mu')}.
\end{align}

The thermal average for each two-qubit Heisenberg operator will be
used to construct the reduced density operator.

\subsection{Reduced density operator and transfer matrix approach}

We will perform the reduced density operator bounded by Ising
particles along a diamond chain. We can proceed by tracing out over
all Heisenberg spins and Ising spins except at the rth block in
Heisenberg spins on a diamond chain. Using the transfer-matrix
approach, we are able to write the reduced density operator by the
following expression
\begin{align}
\rho_{i,j}= & \frac{1}{Z_{N}}\sum_{\{\mu\}}w(\mu_{1},\mu_{2})\ldots w(\mu_{r-1},\mu_{r})\varrho_{i,j}(\mu_{r},\mu_{r+1})\times\nonumber \\
 & w(\mu_{r+1},\mu_{r+2})\ldots w(\mu_{N},\mu_{1}).\label{eq:rho-df}
\end{align}

Alternatively, using the transfer-matrix notation, we can write the
reduced density operator as

\begin{equation}
\rho_{i,j}=\frac{1}{Z_{N}}\mathrm{tr}\left(W^{r-1}P_{i,j}W^{N-r}\right)=\frac{1}{Z_{N}}\mathrm{tr}\left(P_{i,j}W^{N-1}\right),
\end{equation}

where we are assuming

\begin{equation}
P{}_{i,j}=\left[\begin{array}{cc}
\varrho_{i,j}(\tfrac{1}{2},\tfrac{1}{2}) & \varrho_{i,j}(\tfrac{1}{2},-\tfrac{1}{2})\\
\varrho_{i,j}(-\tfrac{1}{2},\tfrac{1}{2}) & \varrho_{i,j}(-\tfrac{1}{2},-\tfrac{1}{2})
\end{array}\right].
\end{equation}

The corresponding matrix $U$ that diagonalizes the transfer matrix
$W$ can be given by
\begin{align}
U=\left[\begin{array}{cc}
\Lambda_{+}-w_{--} & \Lambda_{-}-w_{--}\\
w_{+-} & w_{+-}
\end{array}\right]
\end{align}
and
\begin{align}
U^{-1}=\left[\begin{array}{cc}
\frac{1}{Q} & -\frac{\Lambda_{-}-w_{--}}{Qw_{+-}}\\
-\frac{1}{Q} & \frac{\Lambda_{+}-w_{--}}{Qw_{+-}}
\end{array}\right].
\end{align}

Finally, the reduced density operator defined in eq.(\ref{eq:rho-df})
must be expressed by

\begin{equation}
\rho_{i,j}=\tfrac{\mathrm{tr}\left(U^{-1}P_{ij}U\left[\begin{smallmatrix}\Lambda_{+}^{N-1} & 0\\
0 & \Lambda_{-}^{N-1}
\end{smallmatrix}\right]\right)}{\Lambda_{+}^{N}+\Lambda_{-}^{N}}.
\end{equation}
This result is valid for arbitrary number $N$ of cells in a diamond
chain, under periodic boundary conditions.

\subsection{Reduced density operator in thermodynamic limit}

Real systems are well represented in the thermodynamic limit
($N\rightarrow\infty$), hence, the reduced density operator elements
after some algebraic manipulation becomes

\begin{align}
\rho_{i,j}= & \frac{1}{\Lambda_{+}}\left\{ \tfrac{\varrho_{i,j}(\tfrac{1}{2},\tfrac{1}{2})+\varrho_{i,j}(-\tfrac{1}{2},-\tfrac{1}{2})}{2}+\tfrac{2\varrho_{i,j}(\tfrac{1}{2},-\tfrac{1}{2})w_{+-}}{Q}\right.\nonumber \\
 & \left.+\tfrac{\left(\varrho_{i,j}(\tfrac{1}{2},\tfrac{1}{2})-\varrho_{i,j}(-\tfrac{1}{2},-\tfrac{1}{2})\right)\left(w_{++}-w_{--}\right)}{2Q}\right\} ,\label{eq:rho-elem}
\end{align}
where we have assumed $\left(\Lambda_{-}/\Lambda_{+}\right)^{N}\rightarrow0$
in thermodynamic limit.

All elements of reduced density operator immersed on a diamond chain
are

\begin{equation}
\rho=\left[\begin{array}{cccc}
\rho_{1,1} & 0 & 0 & 0\\
0 & \rho_{2,2} & \rho_{2,3} & 0\\
0 & \rho_{3,2} & \rho_{3,3} & 0\\
0 & 0 & 0 & \rho_{4,4}
\end{array}\right].\label{eq:rho-mat}
\end{equation}

It is worth  noting that, this reduced density operator is the
thermal average two-qubit Heisenberg operator, immersed in the
diamond chain, and it can be verified that ${\rm tr}(\rho)=1$. The cluster approach \cite{ananikian} becomes identical to
transfer-matrix approach (the present approach) only when the magnetic
field is zero. Later, in the next section (Fig. \ref{fig:hvsC}) we
show the difference between both approaches.

\section{Two-qubit Heisenberg entanglement}

Quantum entanglement is a special type of correlation, which only
arises in quantum systems. Entanglement reflects nonlocal
distributions between pairs of particles, even if they are removed
and do not directly interact with each other. In order to measure the
entanglement of anisotropic Heisenberg qubits in the
Ising-Heisenberg model  on a diamond chain, we study the concurrence
(entanglement) of the two-qubits Heisenberg (dimer), which interacts
with two nodal Ising spins using the definition proposed by Wooters
et al. \cite{wootters,hill}.

The concurrence could be  expressed in terms of a matrix
$R$,
\begin{equation}
R=\rho\cdot\left(\sigma^{y}\otimes\sigma^{y}\right)\cdot\rho^{*}\cdot\left(\sigma^{y}\otimes\sigma^{y}\right),
\end{equation}
which is constructed as a function of the density operator $\rho$, given
by Eq. (\ref{eq:rho-mat}), with $\rho^{*}$, which we represent by the complex
conjugate of matrix $\rho$.

Thereafter, the concurrence of two-qubits Heisenberg coupling (bipartite)
could be obtained in terms of eigenvalues of a positive Hermitian matrix $R$:
\begin{equation}
\mathcal{C}(\rho)=\mathrm{max}\{\sqrt{\lambda_{1}}-\sqrt{\lambda_{2}}-\sqrt{\lambda_{3}}-\sqrt{\lambda_{4}},0\},\label{eq:Cdf}
\end{equation}
with eigenvalues $\lambda_{1}\geqslant\lambda_{2}\geqslant\lambda_{3}\geqslant\lambda_{4}$.
Equation (\ref{eq:Cdf}) can be reduced to

\begin{equation}
\mathcal{C}(\rho)=2\mathrm{max}\{\rho_{2,3}-\sqrt{\rho_{1,1}\rho_{4,4}},0\}.\label{eq:conc-def}
\end{equation}

\subsection{Concurrence}

As presented above the entanglement can be studied in terms of concurrence
defined by Eq. (\ref{eq:conc-def}), as a function of  Hamiltonian
parameters defined in Eq. (\ref{eq:Hamil-1}), as well as temperature
and external magnetic field.

\begin{figure}
\includegraphics[scale=0.3]{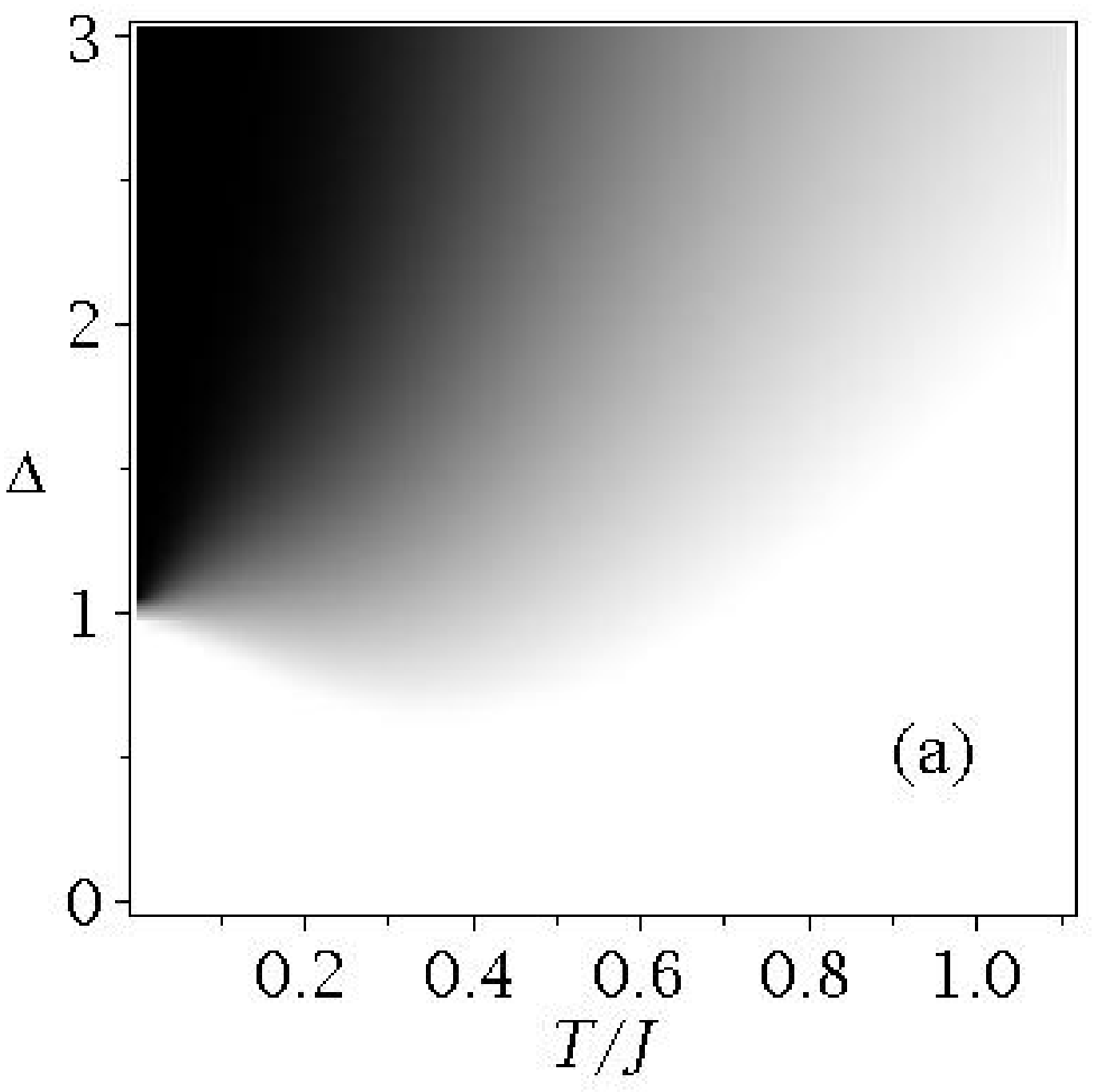}\includegraphics[scale=0.3]{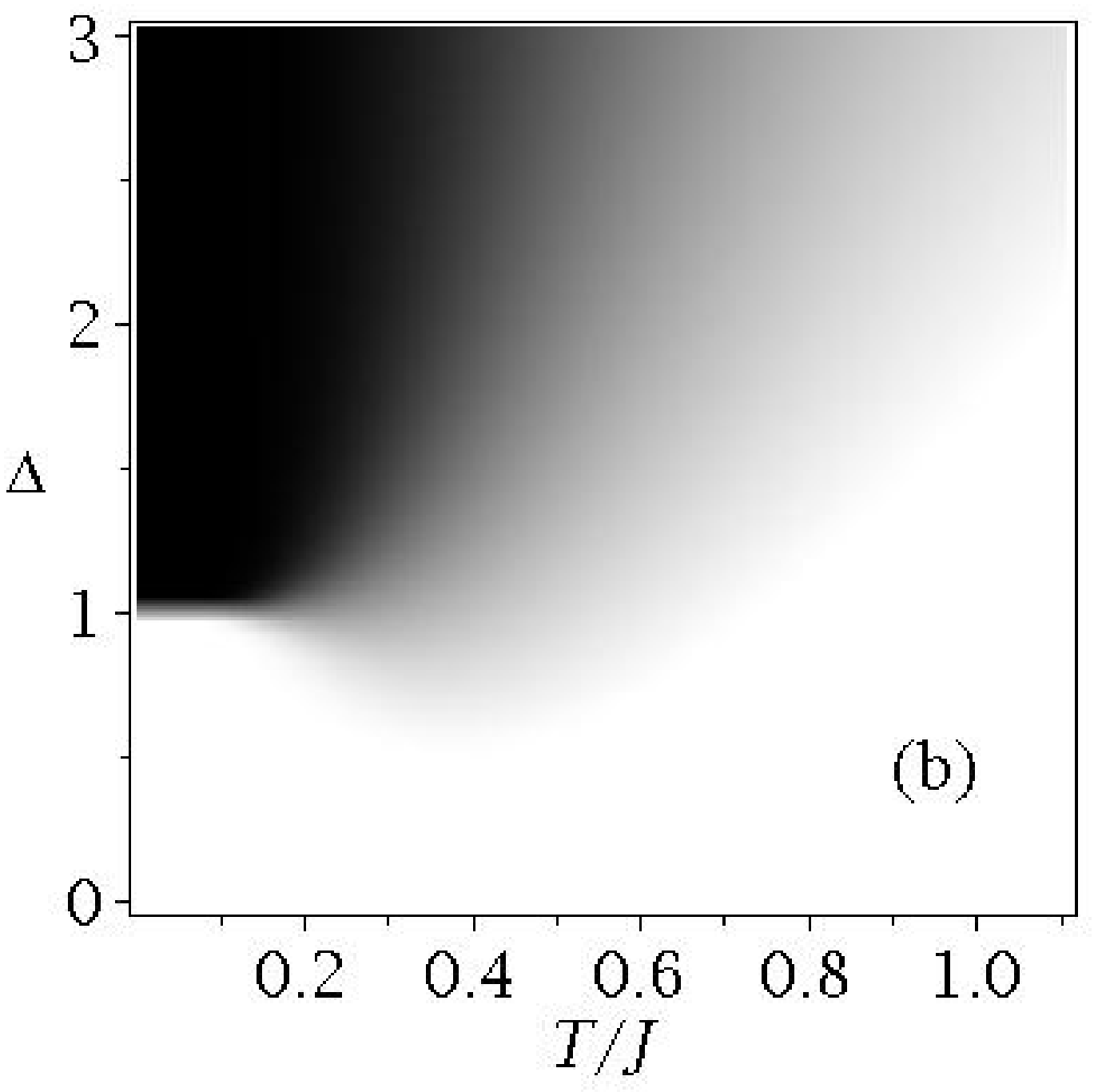}\caption{\label{fig:TvsDlt}Density plot of  concurrence $\mathcal{C}$ as function of $T/J$
and $\Delta$.  Black color corresponds to the maximally entangled
region, while by white color is the unentangled region,
the gray region  means  the entanglement with  degrees of different intensity.
In (a) we display $\mathcal{C}$ for absence of the magnetic field and in
(b) we display $\mathcal{C}$ for $h/J=1$.}
\end{figure}

In Fig. \ref{fig:TvsDlt} we illustrate the density plot of
concurrence  $\mathcal{C}$ as a function of $T/J$ and $\Delta$ for a
fixed value of  $J_{1}/J$=1. Black corresponds to the maximum entangled region ($\mathcal{C}=1$), while white 
is the unentangled region  ($\mathcal{C}=0$). Gray means the different degrees of
entanglement  ($0<\mathcal{C}<1$). After that we  use this representation for the
concurrence $\mathcal{C}$, depending on the parameters of the
Hamiltonian \eqref{eq:Hamil-1}.  The entangled region represented by different gray
intensities depends also on the temperature. For high temperature the
fuzzy region increases, while for the low-temperature phase between the
entangled region and the unentangled region the boundary becomes
sharper. In Fig. \ref{fig:TvsDlt}(a) it is shown that the model is
maximally entangled only for $\Delta\geqslant1$ in the absence of
the magnetic field, while  the concurrence $\mathcal{C}$ is always
 less than 1 for $\Delta<1$ . The concurrence  $\mathcal{C }$
becomes smaller with increasing temperature and the entanglement disappears at high temperatures. In Fig. \ref{fig:TvsDlt}(b) we display $\mathcal{C}$ for $h/J=1$, where
the concurrence behaves similar to Fig. \ref{fig:TvsDlt}(a). But there is the maximum entangled region Fig. \ref{fig:TvsDlt}(b) for  $\Delta\geqslant1.0$ and at temperatures less than $T/J\thickapprox0.2$.

\begin{figure}
\includegraphics[scale=0.3]{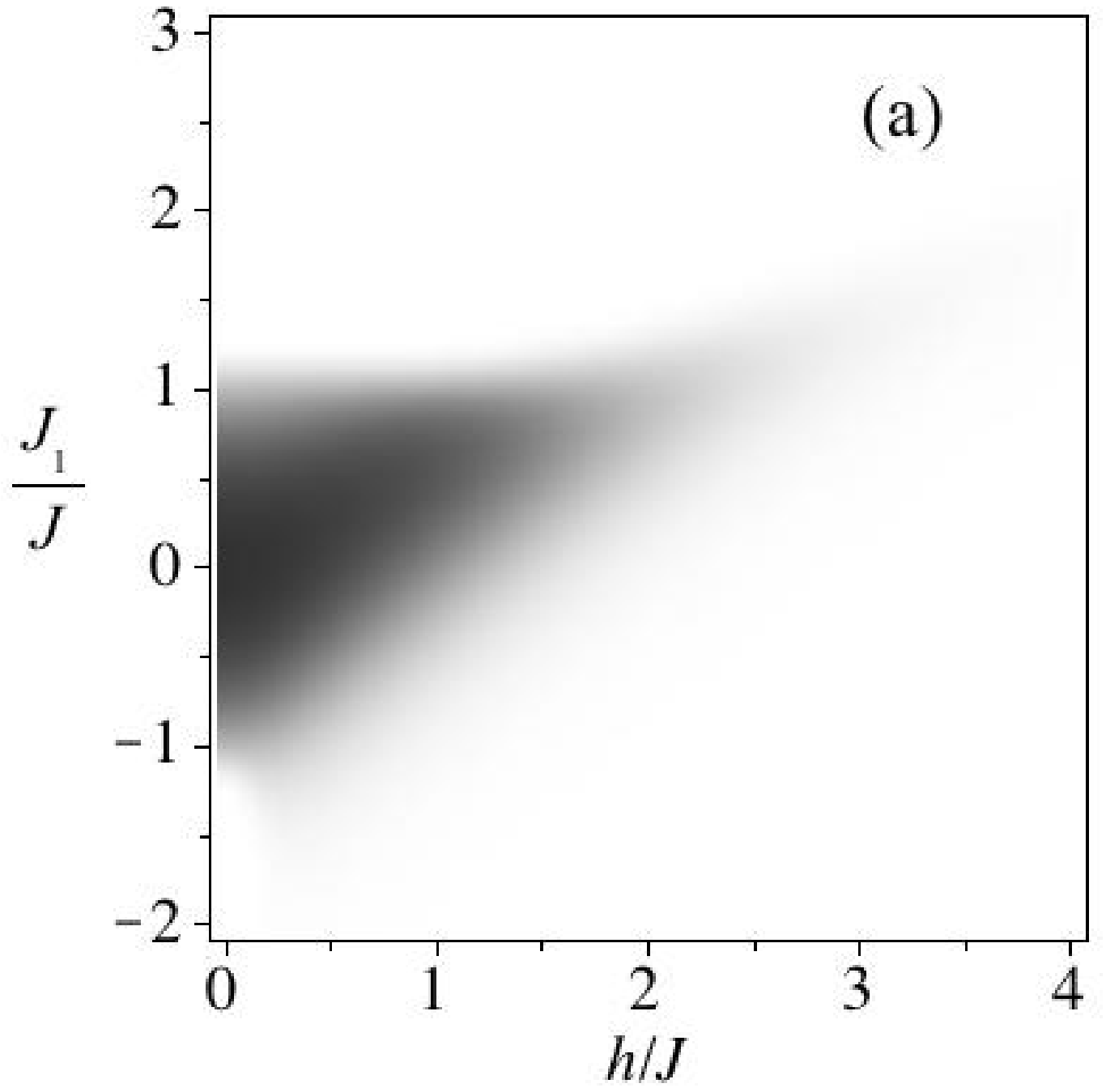}\includegraphics[scale=0.3]{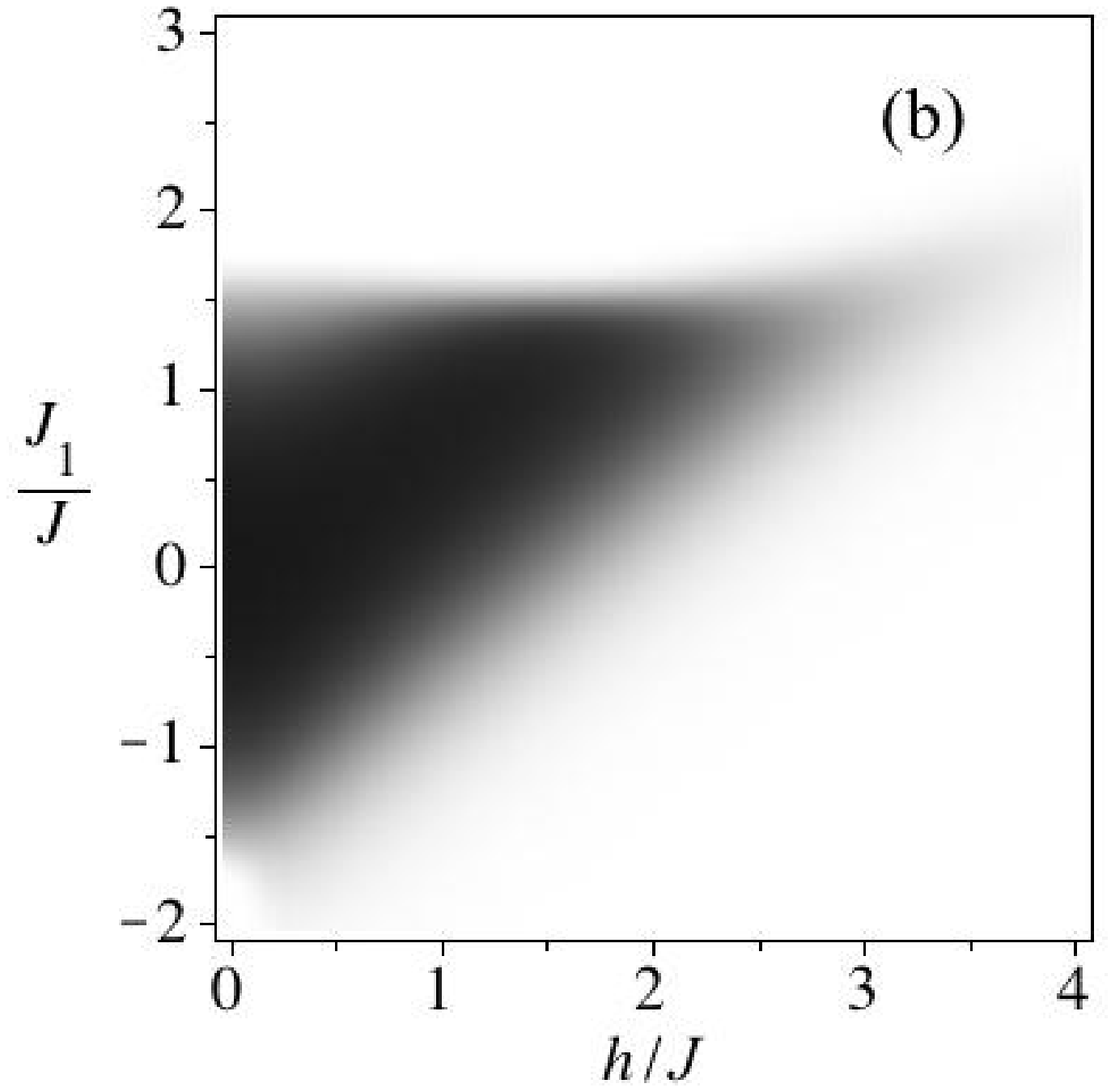}

\caption{\label{fig:hvsJ1}Density plot of  concurrence $\mathcal{C}$ as function of $h/J$
and  $J_{1}/J$ . In (a) we display of
   $\Delta=1.0$ and  (b) is for $\Delta=2.0$. The black (white) region corresponds to $\mathcal{C}=1 (0)$, and by gray regions we indicate a concurrence $0<\mathcal{C}<1$.}
\end{figure}

The density plot of  concurrence  $\mathcal{C}$ as a function of magnetic field $h/J$ and $J_{1}/J$ is shown in FIG.\ref{fig:hvsJ1}  . To represent
the concurrence we use the same representation as in Fig. \ref{fig:TvsDlt},
for a fixed value of $T/J=0.3$. In Fig. \ref{fig:hvsJ1}(a) we display
$\mathcal{C}$ for a $\Delta=1.0$; thus, we can show that the concurrence
is always less than  $\mathcal{C}\lesssim 0.5$. The largest entanglement occurs for small $|J_{1}/J|\lesssim1$
and small magnetic field $h/J\lesssim2$. Whereas in Fig. \ref{fig:hvsJ1}(b)
we display $\mathcal{C}$ for $\Delta=2.0$, and the entanglement becomes stronger
than for $\Delta=1$, but the concurrence is still limited to the
regions $|J_{1}/J|\lesssim1.5$ and $h/J\lesssim3$.

\begin{figure}
\includegraphics[scale=0.3]{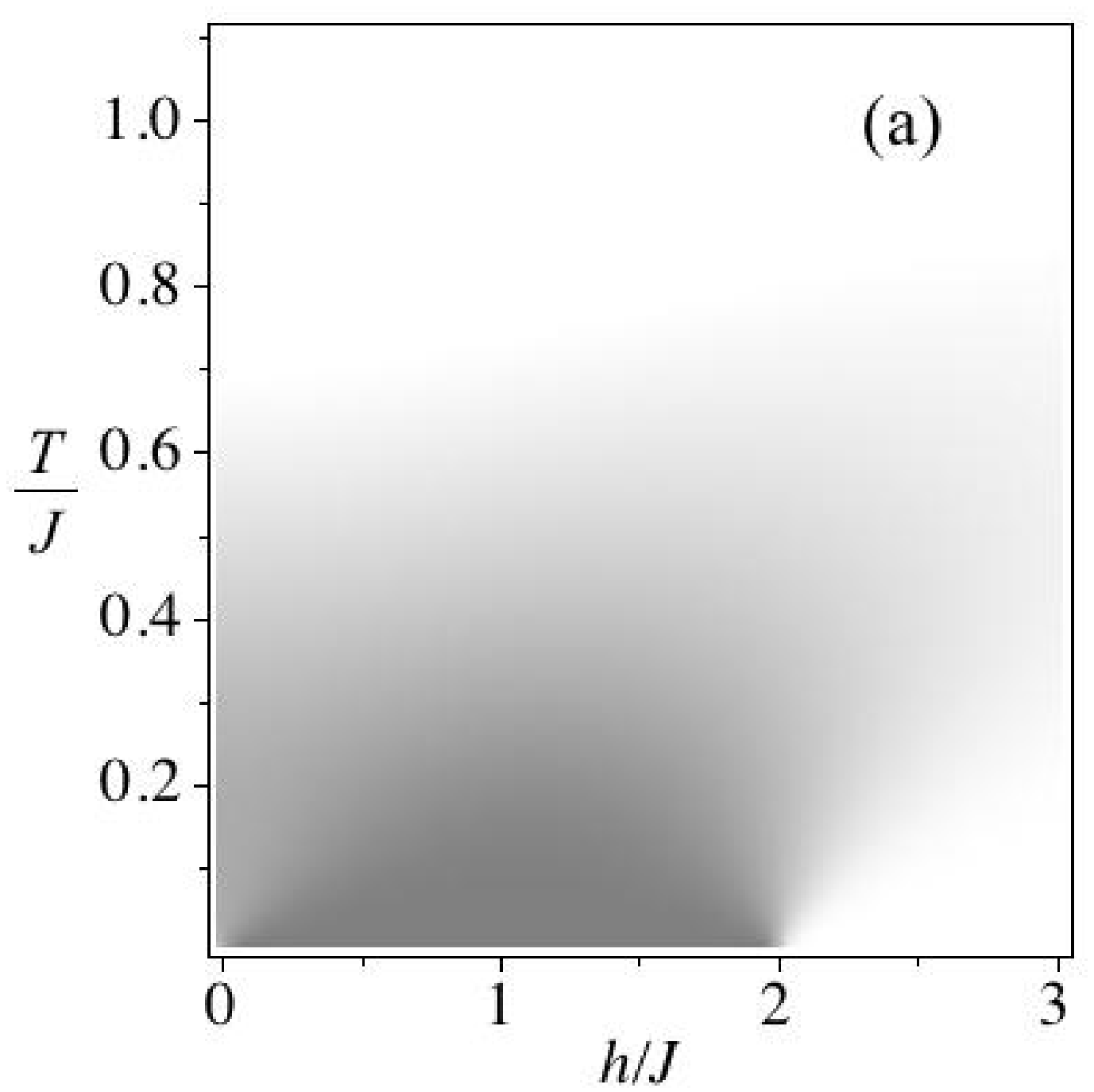}\includegraphics[scale=0.3]{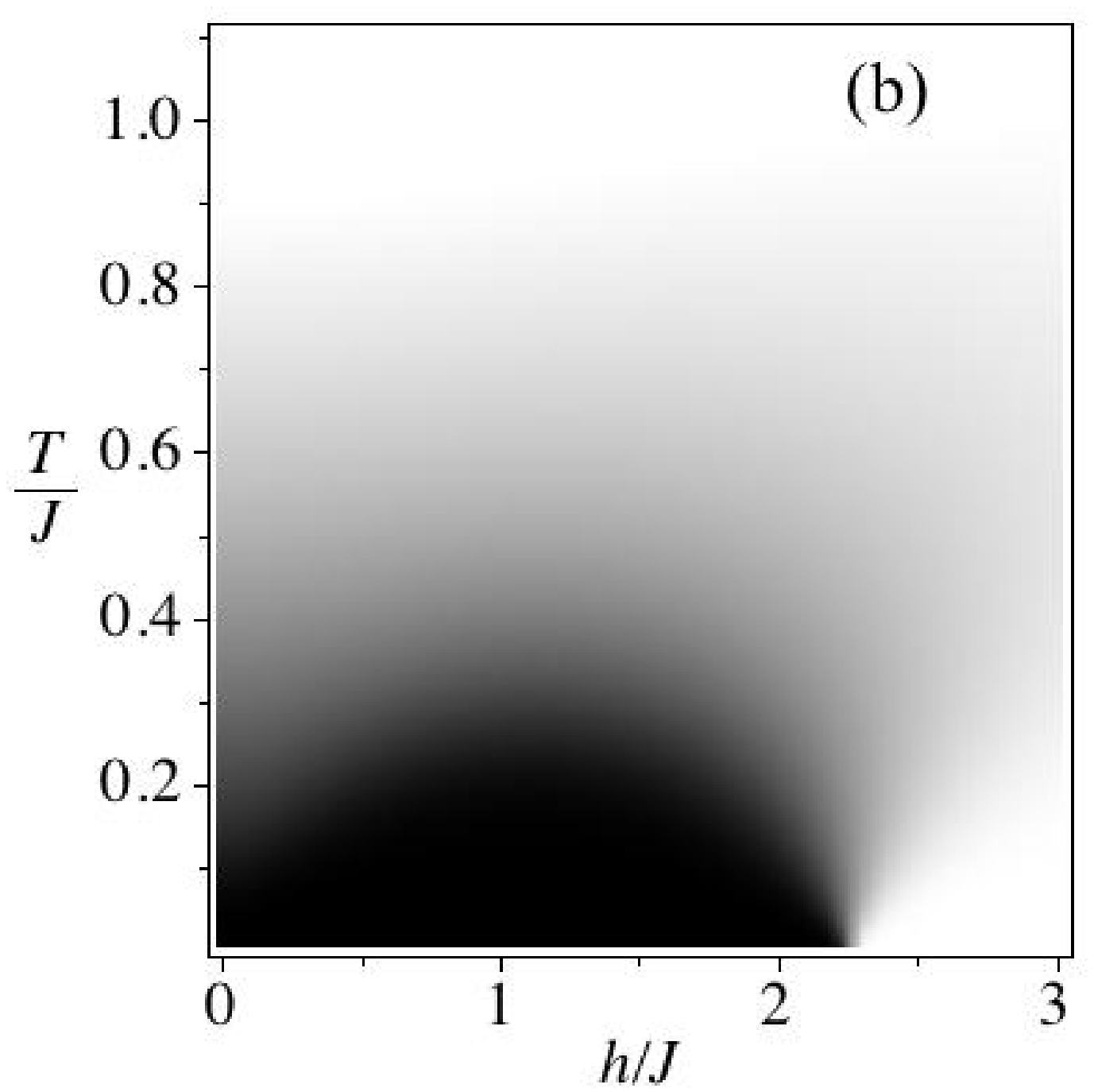}\caption{\label{fig:hvsT}Density plot of concurrence
$\mathcal{C}$ as a function of $h/J$ and $T/J$:  In (a) we display
$\mathcal{C}$ for a $\Delta=1$ and in (b) we display $\mathcal{C}$ for
$\Delta=1.5$. The black (white) region corresponds to $\mathcal{C}=1 (0)$, and by gray regions we indicate a concurrence $0<\mathcal{C}<1$.}
\end{figure}

Another  case is shown in Fig. \ref{fig:hvsT} for the
density plot of concurrence  $\mathcal{C}$ as a function of $h/J$
and $T/J$, at a fixed value of $J_{1}/J=1$. Black (white) region indicates the entangled (unentangled) region, while the gray region corresponds to $0<\mathcal{C}<1$. The
concurrence is not very significant  for $\Delta=1$, as shown in  Fig.
\ref{fig:hvsT}(a). For values of $0\leqslant h/J\lesssim2$ the
entanglement becomes $\mathcal{C}\lesssim0.25$ at small
temperatures, while at higher temperatures it disappears. For
$\Delta=1.5$ the situation has changed; we observe the maximally
entangled region for values of $0\leqslant h/J\lesssim2.3$ and
$T/J\lesssim0.2$, as shown in Fig. 5 (b), and at higher
temperatures and a strong magnetic field entanglement vanishes asymptotically.

\begin{figure}
\includegraphics[scale=0.218]{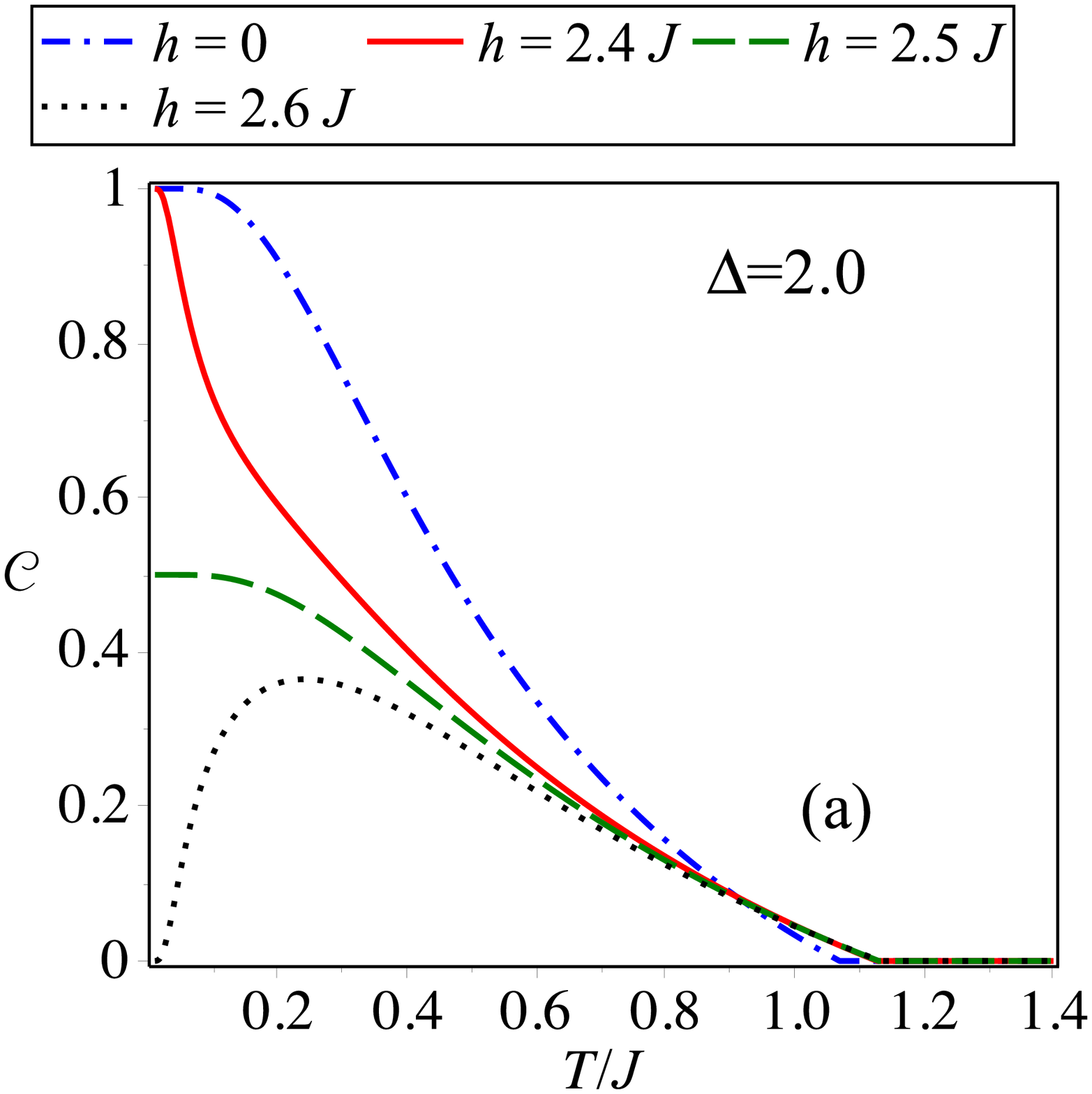} \includegraphics[scale=0.218]{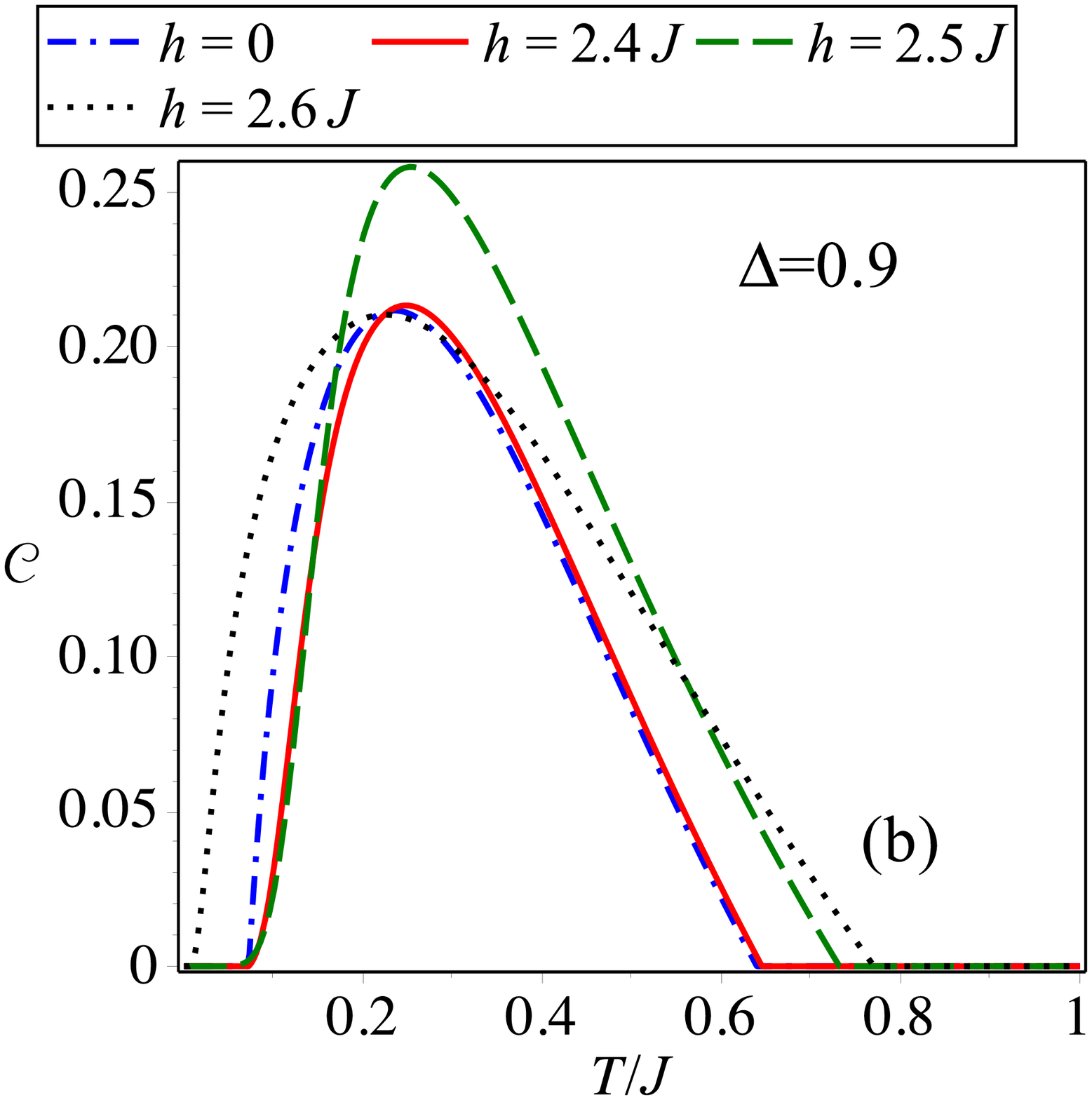}
\caption{\label{fig:TvsC} (Color online) Concurrence as a function of temperature for a fixed
value of $J_{1}/J$=1: In (a) we displays concorrence for $\Delta=2$ and (b) displays concorrence for $\Delta=0.9$.}

\end{figure}

The concurrence as a function of  temperature $T/J$ for fixed values
of $J_{1}/J=1$ is shown in Fig. \ref{fig:TvsC}, for several values of
the magnetic field. First of all, we illustrate the concurrence  for
$\Delta=2$ in Fig. \ref{fig:TvsC}(a). The concurrence reaches its
maximum value $\mathcal{C}=1$ in the absence of magnetic field and
at low temperature, whereas at high temperature entanglement
disappears,  the threshold temperature occurs at  $T/J\approx1.1$,  which is represented by a dash-dotted  line. The
behavior of the concurrence varies in a strong magnetic field, and the
entanglement curve in most cases is lower than in the absence of a
magnetic field. In the case of $h/J=2.4$  (solid line), the maximum value of
concurrence  $\mathcal{C}=1$ is only achieved at very low temperature  $T/J$,
although the threshold temperature $T/J$ is slightly larger than that in
the absence of a magnetic field. In the case of the magnetic field be
$h/J=2.5$ (dashed  line), the concurrence achieves  the highest value  around  $\mathcal{C}=0.5$
at zero temperature. While for $h/J=2.6$ (dotted line), the system becomes
unentangled at zero temperature, but at low temperatures near
$T/J\thickapprox0.3$ it reaches its maximum value  around
$\mathcal{C}\thickapprox0.38$.
Second, in Fig. \ref{fig:TvsC}(b)
 the behavior of the concurrence for a fixed value of the anisotropic parameter $\Delta=0.9$, and for  different magnetic field
 values are shown. Now, the behavior of  concurrence decreases with different
values of the magnetic field and has a peak near
$T/J\thickapprox0.3$. The concurrence for all of these magnetic fields are unentangled at zero temperature and there is
a temperature threshold near to $T_{th}/J\thickapprox0.64$ for $0\leqslant h/J\lesssim  2.4$, while for $h/J\gtrsim2.5$ the threshold temperature jumps to  $T_{th}/J\gtrsim0.75$.

\begin{figure}
\includegraphics[scale=0.218]{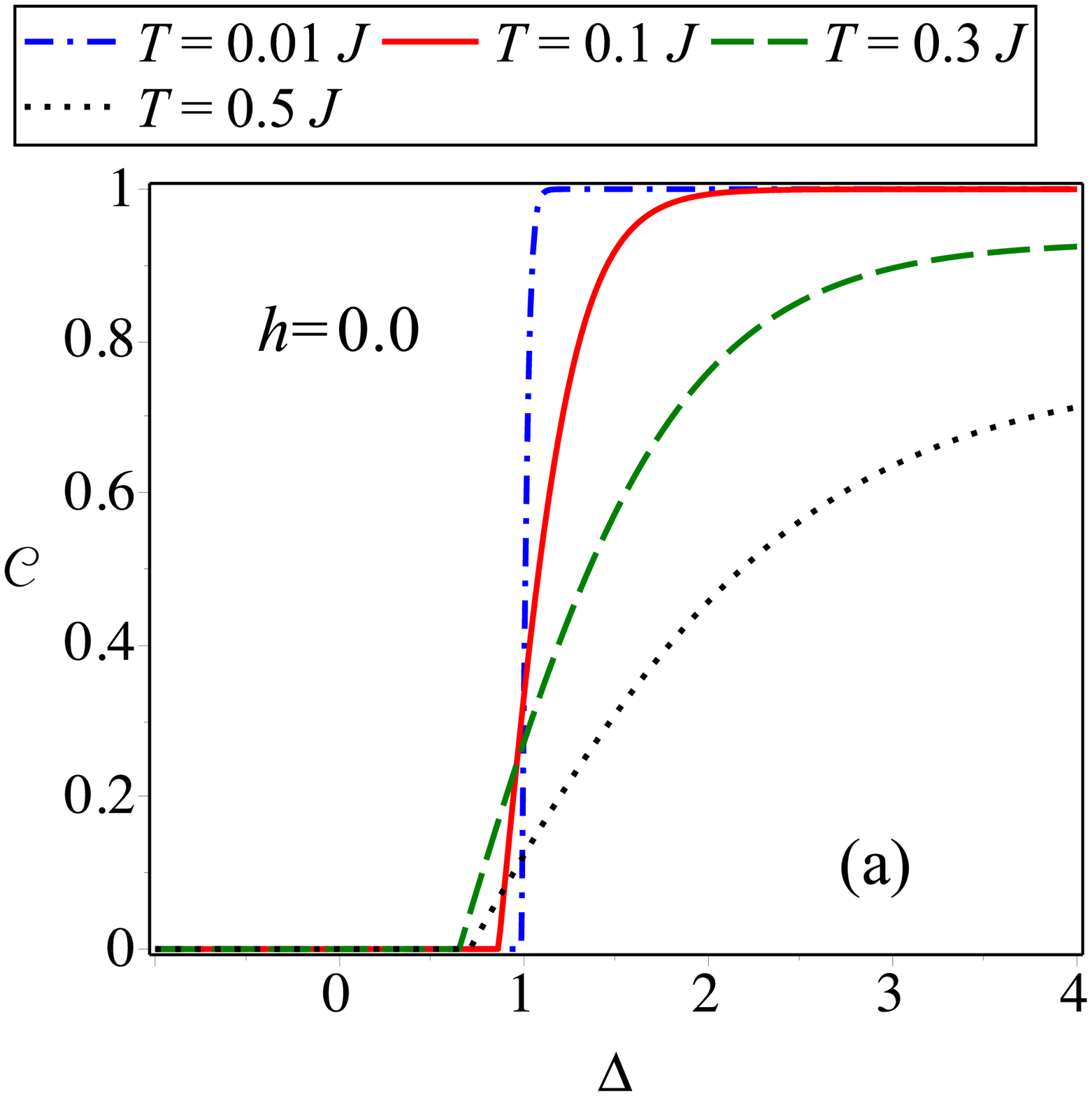}\includegraphics[scale=0.218]{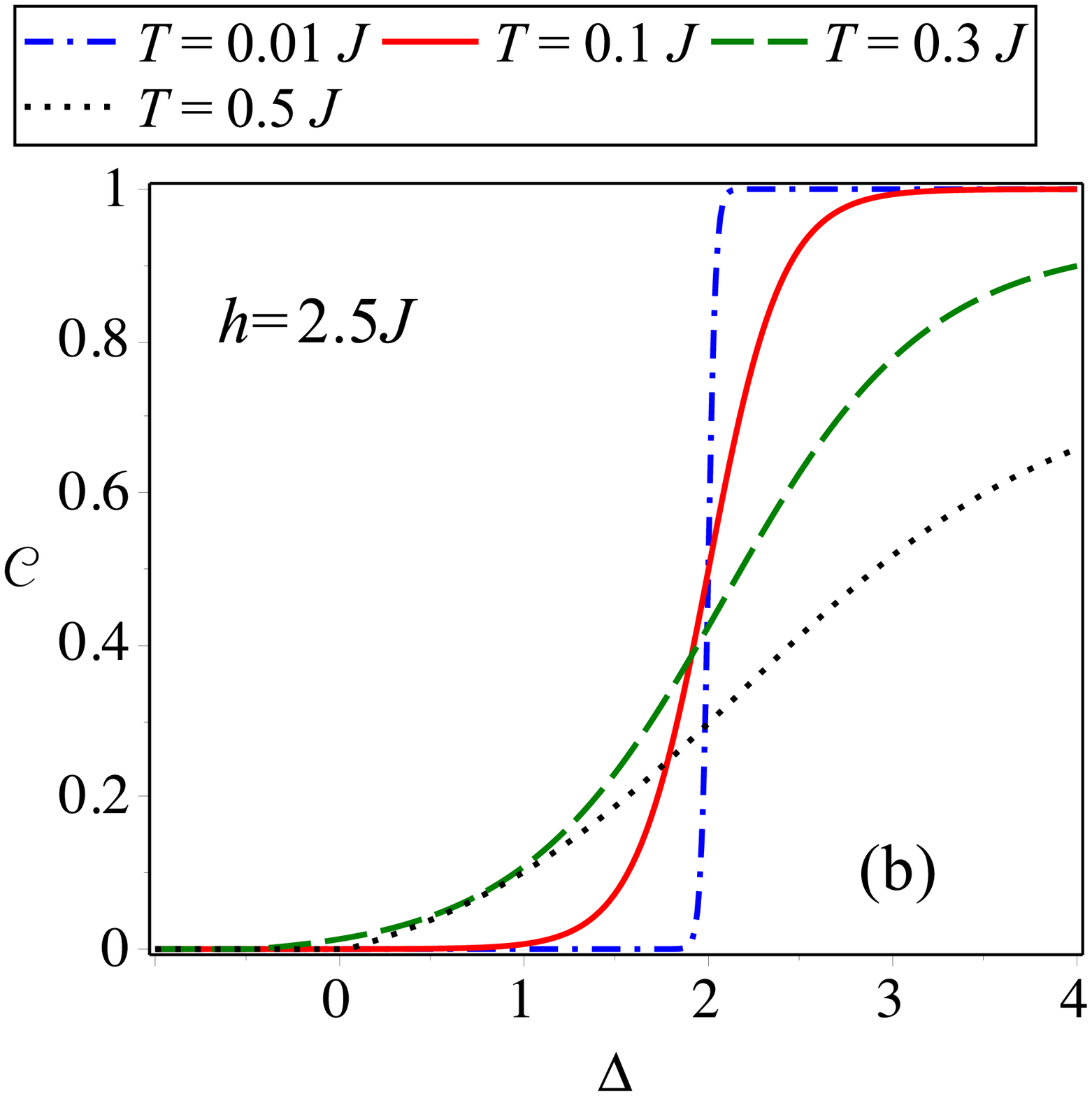}\caption{\label{fig:DltvsC}(Color online) Concurrence as a function of anisotropic parameter
$\Delta$:(a) $h/J=0$ and (b) $h/J=2.5$ with different fixed values of temperature.}

\end{figure}

In  Fig. \ref{fig:DltvsC}(a)  the concurrence as a function
of the anisotropy parameter $\Delta$ in the absence of a magnetic
field for various fixed temperatures is plotted. For $\Delta\lesssim 1$ the system is
unentangled, and  for $\Delta\gtrsim1$ the system becomes entangled. The
maximum of the entangled region is achieved for larger anisotropic
parameters, but when the temperature increases the entanglement decreases. The concurrence as a
function of the anisotropy parameter $\Delta$, and  for  a fixed magnetic
field $h/J = 2.5$ is plotted in Fig. \ref{fig:DltvsC}(b), for the same
set of temperatures as shown in Fig. \ref{fig:DltvsC}(a). Here we can observe the influence of the magnetic field for the concurrence; at
low temperature ($T/J=0.01$) the system is entangled for the
values around of $\Delta\thickapprox2$, whereas for higher temperature  a small entanglement appears even for $\Delta<1$.

\begin{figure}
\includegraphics[scale=0.225]{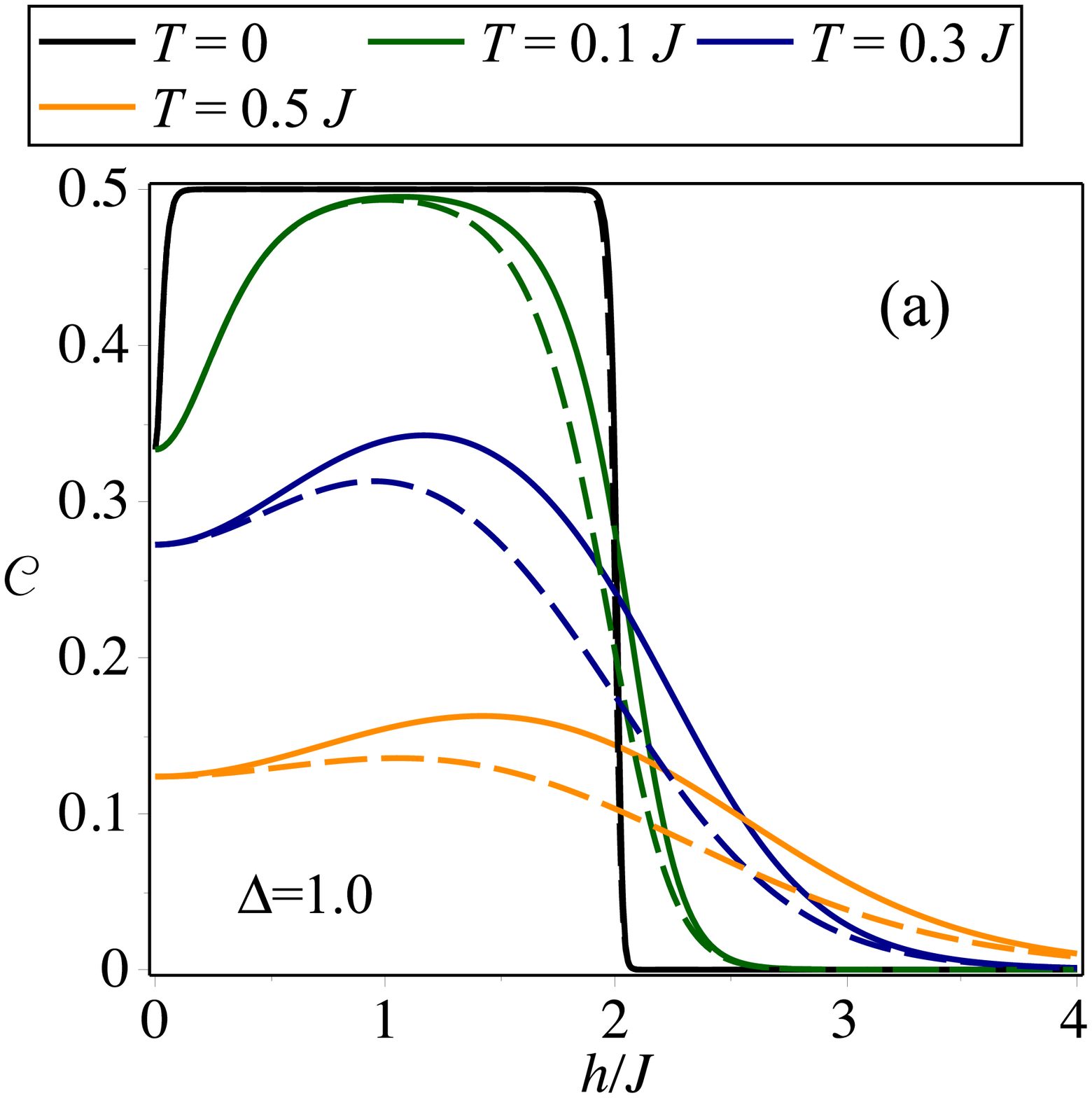}\includegraphics[scale=0.225]{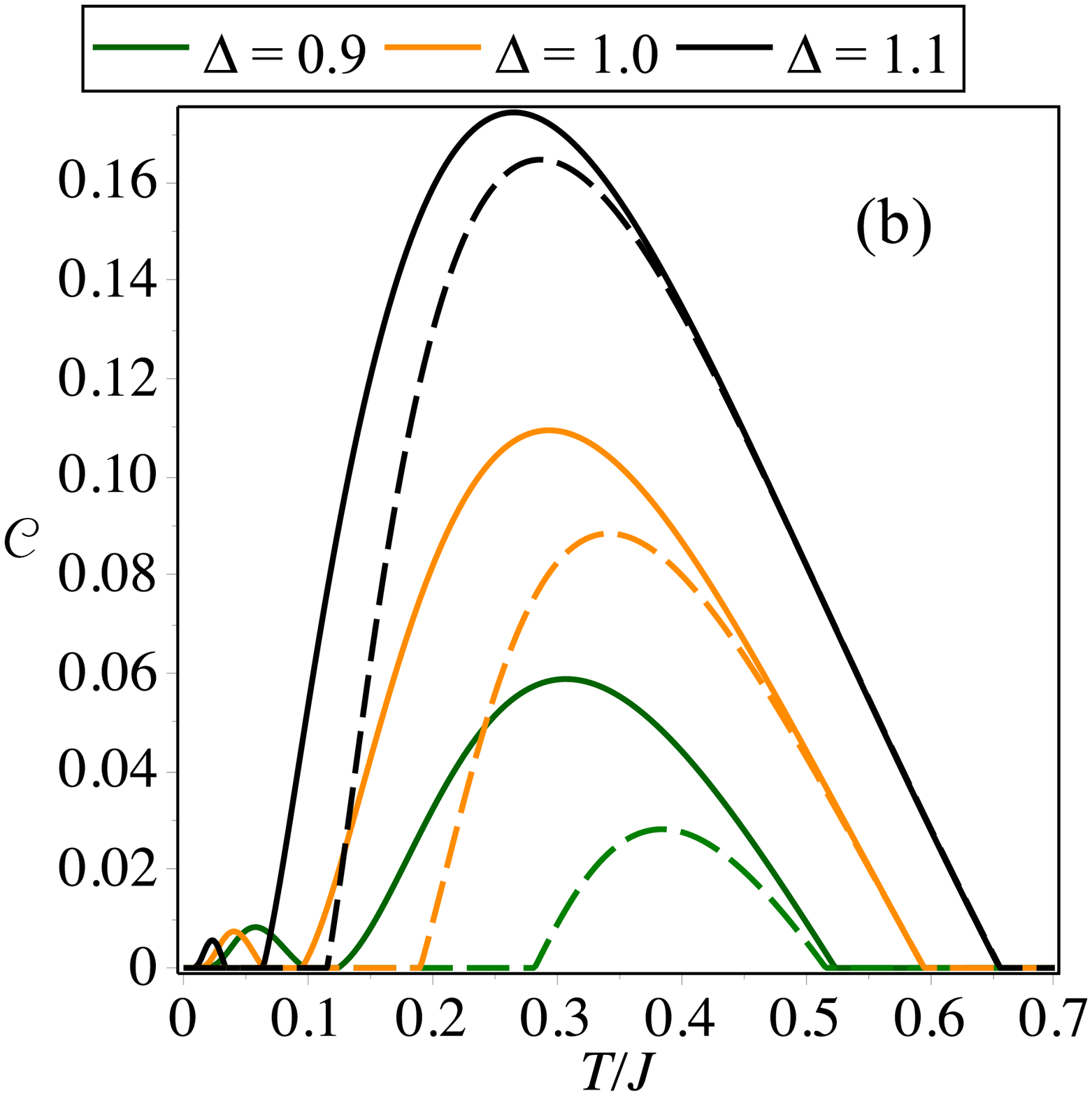}
\caption{\label{fig:hvsC} (Color online) (a) Concurrence against external magnetic field, for
different values of temperature with anisotropic parameter
$\Delta=1$. (b) Concurrence as a function of temperature assuming fixed values of the anisotropic parameter and a fixed magnetic field $hJ=0.2$. By the solid lines we represent the cluster-approach results \cite{ananikian} whereas by the dashed lines we represent our result.}
\end{figure}

Finally, in Fig. \ref{fig:hvsC}(a) we display the concurrence as a function
magnetic field, assuming a fixed value of the anisotropy parameter $\Delta=1$.
In this plot  the difference between the cluster
approach \cite{ananikian}  and  the transfer-matrix approach used here is also shown.
 By the solid lines we represent the transfer-matrix
approach, while with dashed lines we denote the cluster approach \cite{ananikian}.
While, in Fig. \ref{fig:hvsC}(b)  the concurrence as a function of temperature ($T/J$) is illustrated, by the continuous line we observe the cluster approach, and by the dashed line we represent our current result. For the anisotropic parameter $\Delta=1.1$ and for fixed  magnetic field $h/J=0.2$.  Clearly, the cluster approach shows a double peak (solid line), whereas by dashed lines we represent our current approach discussed here.  Therefore, we can conclude that the small peak is absorbed due to the Ising coupling of the diamond chain structure.
However, the small peak is rather irrelevant since the concurrence maximum value is around $\mathcal{C}\thickapprox 0.01$.

 The difference between both approaches is consistent, since the
entanglement on the diamond chain must be lower than for the cluster
approach, where only the nearest Ising spins are considered, and
we ignored the remaining diamond blocks coupling contributions, so the cluster approach is equivalent to infinite uncoupled diamond blocks.

\subsection{Entanglement  threshold temperature}

\begin{figure}
\includegraphics[scale=0.218]{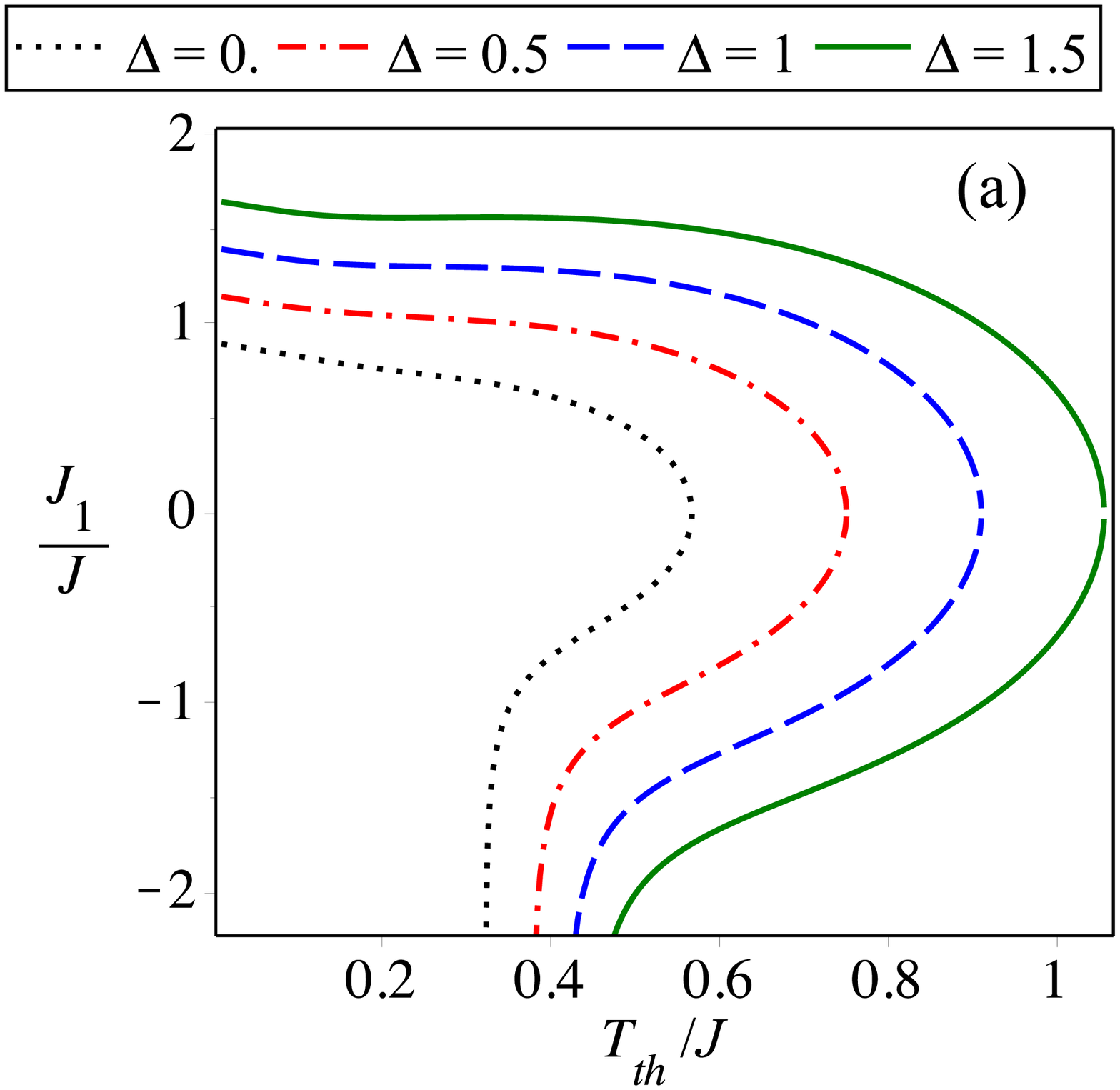} \includegraphics[scale=0.218]{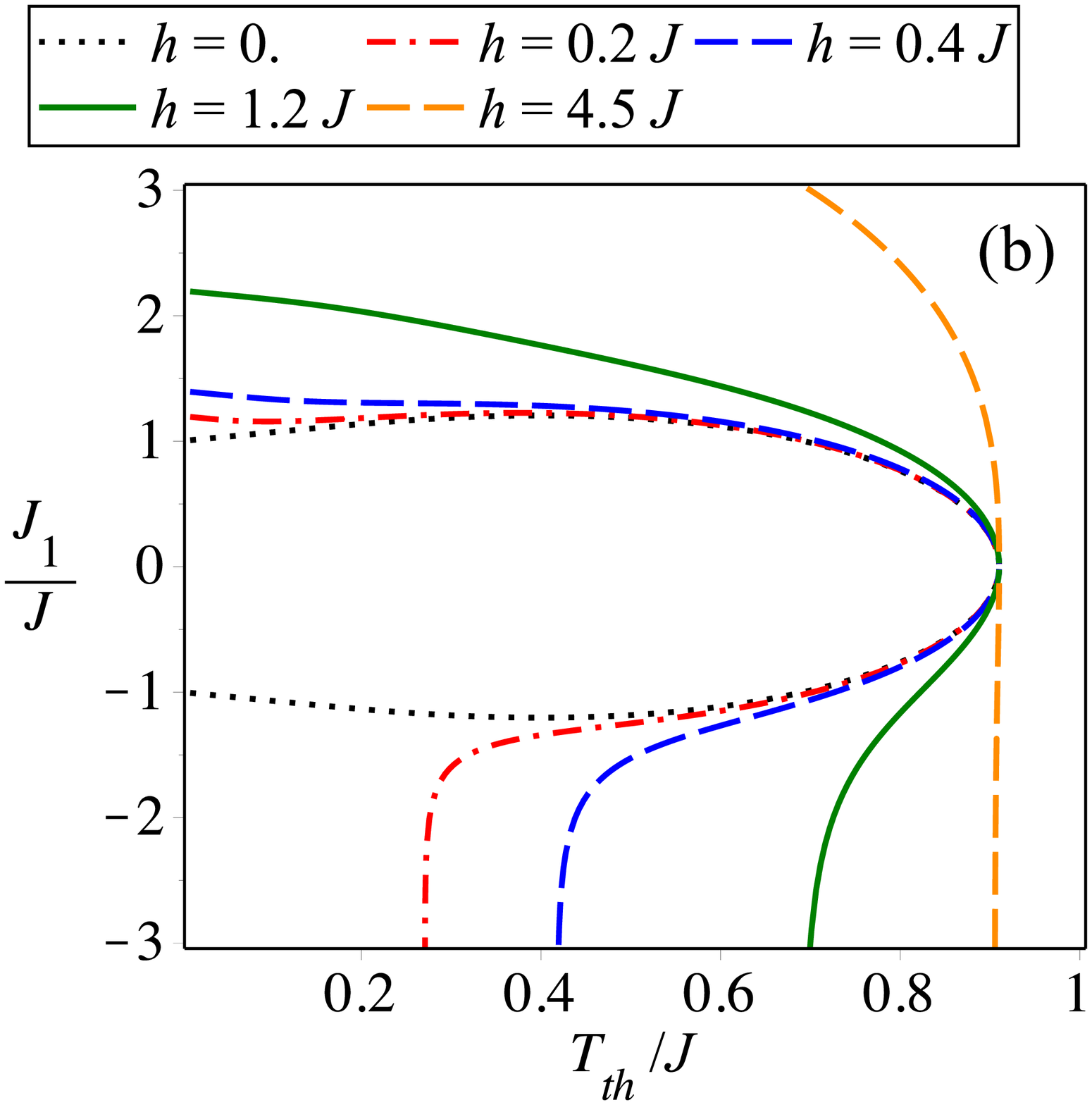}

\caption{\label{fig:PhJ1Td}(Color online) Phase diagram, of Ising coupling parameter $J_{1}/J$
as a function threshold temperature $T_{th}/J$. (a) For several values
of anisotropy parameter $\Delta$. (b) For several values of external
magnetic field $h/J$.}
\end{figure}

In Fig. \ref{fig:PhJ1Td}(a) we display the phase diagram of the entangled region and the untangled region, as a function of the Ising coupling parameter
$J_{1}/J$ against threshold temperature for a fixed values of magnetic
field $h/J=0.1$, and for several anisotropy parameters. For each
curve the left side ($T/J<T_{th}/J)$ of this region corresponds to
the entangled state, while for right side ($T/J>T_{th}/J$) the system
becomes an unentangled region. The entangled region
depends of the anisotropy parameter $\Delta$; as long as the anisotropy
parameter increases the entangled region increases, too. For a large
negative value of $J_{1}/J<0$, the unentangled region is limited
around $T_{th}/J\gtrsim0.3$, whereas, for $T_{th}/J\lesssim0.3$
we still have an entangled region, although this entanglement leads
to a weakly entangled region for large negative value of  $J_{1}/J$, even for
small a anisotropy parameter $\Delta$. However, for large positive value of $J_{1}/J$
the system becomes a fully  unentangled region, but for large values of anisotropy
parameter $\Delta$ tries to enhance the entanglement, although it always
becomes an unentangled region for  $J_{1}/J$ large enough.
In Fig. \ref{fig:PhJ1Td}(b) we display the phase diagram of entangled region and untangled region as a function of the Ising coupling parameter $J_{1}/J$ against threshold temperature for a fixed value
of the anisotropy parameter $\Delta=1$, and for several values of magnetic
field. For null magnetic field the entangled region is limited by
the dotted curve, (the upper and lower side of this curve are symmetric),
but as soon as the magnetic field is turned on, this symmetry is broken
even for a low magnetic field like $h/J=0.2$ (described by the dash-dotted line), the system will
 always be entangled for $J_{1}/J\lesssim -1.5$ and $T_{th}/J\lesssim0.28$.
For a stronger magnetic field this limit increases until it achieves the asymptotic
limit $T_{th}/J\approx0.910239$. The largest value of threshold temperature
$T_{th}/J\approx0.910239$ occurs when $J_{1}/J=0$ and for arbitrary
values of magnetic field. Of course, this limiting threshold temperature
also depends on the magnitude of anisotropic parameters, as we can verify
in Fig. \ref{fig:PhJ1Td}(a).

\begin{figure}
\includegraphics[scale=0.3]{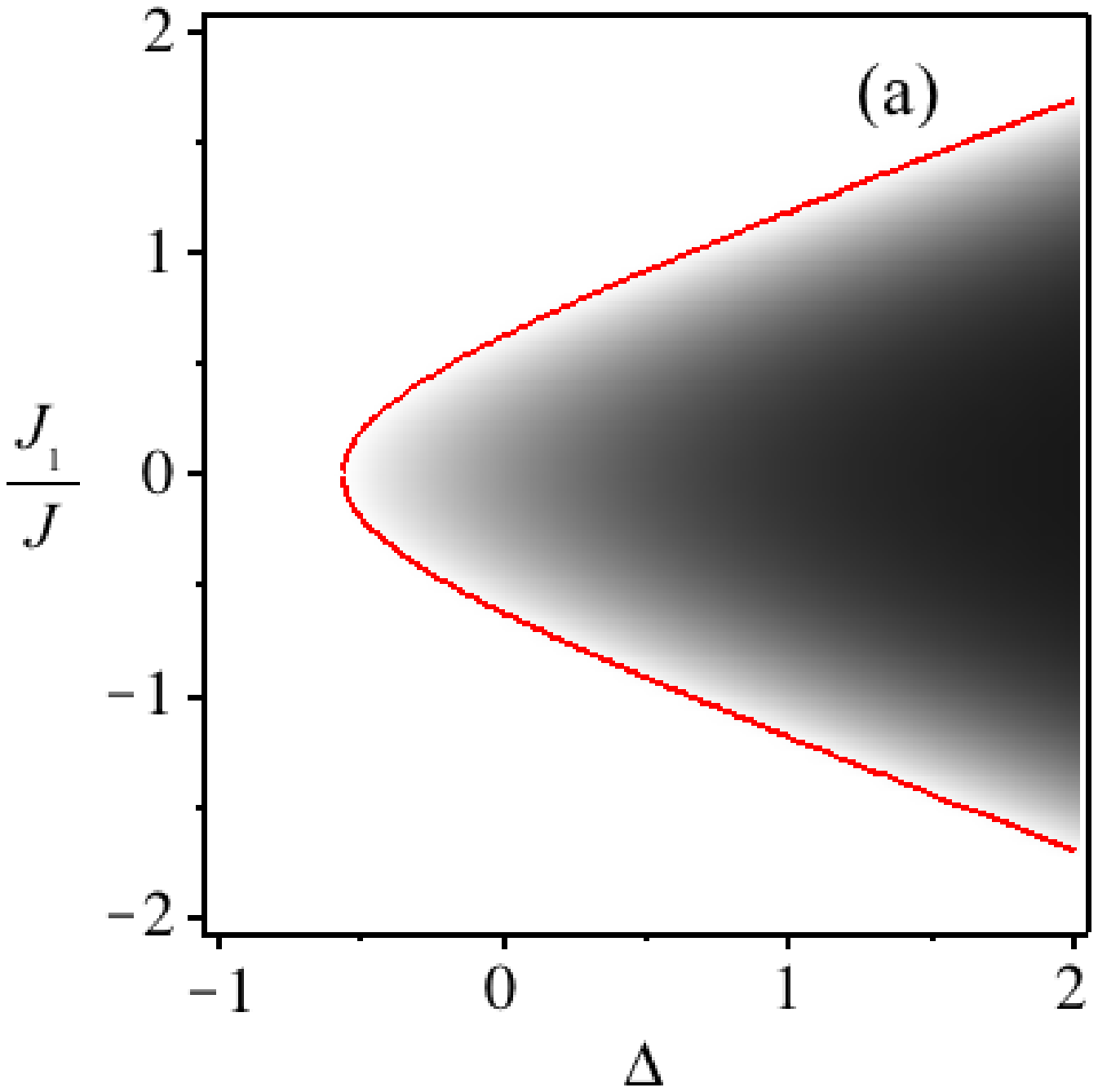}\includegraphics[scale=0.3]{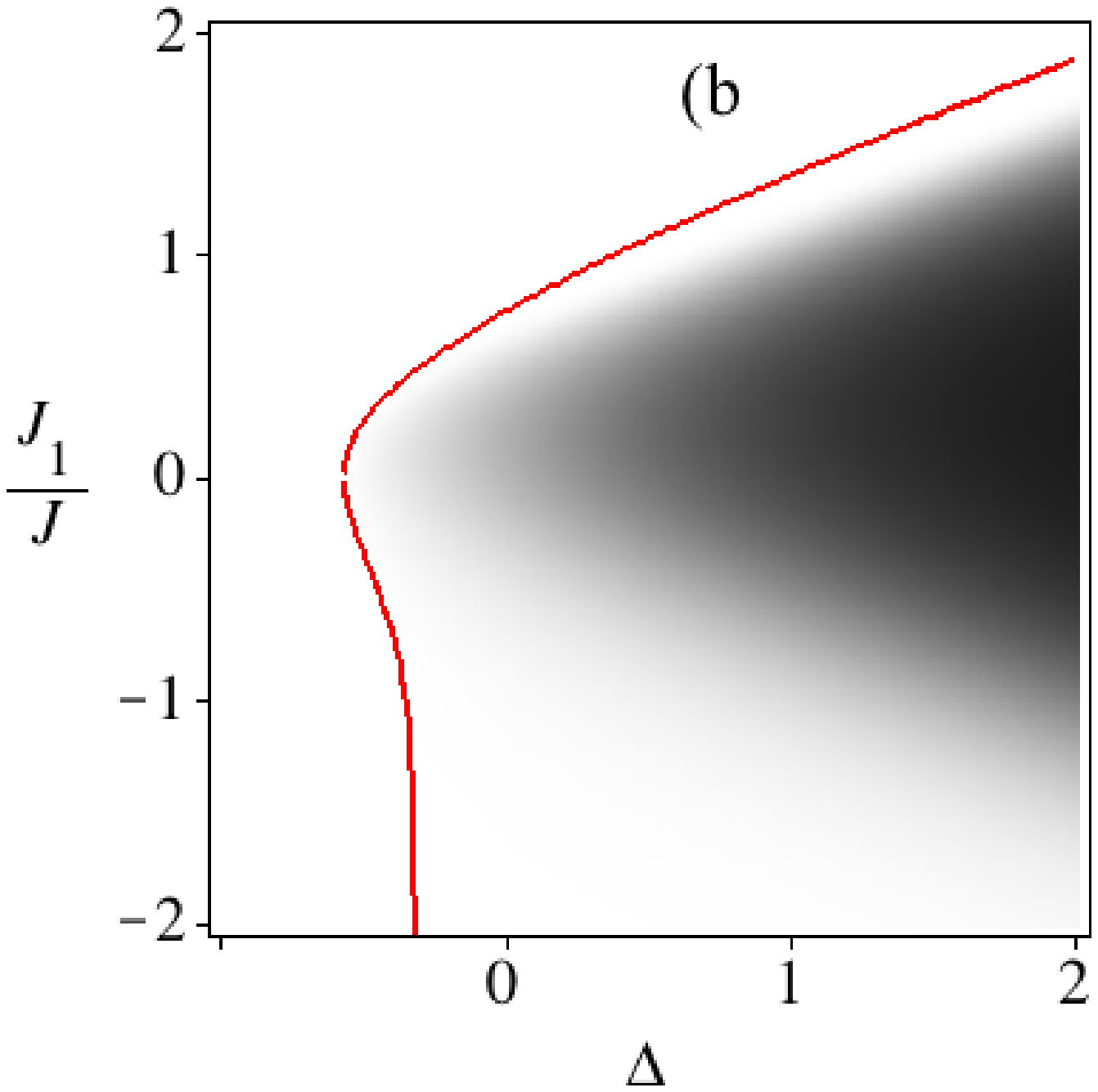}\caption{\label{fig:DltvsJ1}(Color online) Density plot concurrence $\mathcal{C}$ as function of
$\Delta$ versus $J_{1}/J$. In (a) we display $\mathcal{C}$
for a null magnetic field and in (b), we display $\mathcal{C}$ $\mathcal{C}$ for $h/J=1$.}
\end{figure}

Other important density plots, are displayed in Fig. \ref{fig:DltvsJ1}, for
the concurrence $\mathcal{C}$ as a function of $\Delta$ against
$J_{1}/J$, assuming a fixed temperature $T/J=0.3$.  This density plots follow same pattern (definition) as those illustrated in Fig. \ref{fig:TvsDlt}; the black (white) region corresponds to  $\mathcal{C}=1  (0)$, while a gray region corresponds to $0<\mathcal{C}<1$. In
Fig. \ref{fig:DltvsJ1}(a) the maximally entangled region only appears
for $\Delta>1$; we can also verify that the illustration is symmetric
[($\mathcal{C}(J_{1}/J)=\mathcal{C}(-J_{1}/J))$] only when $h=0$.
However, when the external magnetic field is turned on [i.e., ($h/J=1$)], the
symmetry $J_{1}/J\leftrightarrow-J_{1}/J$ is broken as shown in Fig.
\ref{fig:DltvsJ1}(b). When $J_{1}/J>0$ the boundary between the entangled region
and the unentangled region is sharper, while for $J_{1}/J<0$ the boundary
between the entangled region and the unentangled region becomes less pronounced.
The maximally entangled region becomes stronger  for a large
value of the anisotropy parameter $\Delta>1$ compared to the entanglement
without magnetic field. This result is similar to the previous result
discussed by Zhou et al. \cite{Zhou}, when they studied the anisotropy
effect of a two-qubit Heisenberg model with external magnetic field,
although they discussed only one-couple qubits.
The red solid line is used to represent the boundary between the entangled region
and the unentangled region. This curve is also known as the threshold-temperature
curve; clearly for null magnetic field the threshold temperature rounds
symmetrically the entangled region-this is not so evident when external magnetic
field is switched on. Even for a rather weak external magnetic field,
the entangled region spreads out for $J_{1}/J<0$. The entanglement emerge for  $J_1/J<-1$,  [Fig.  \ref{fig:DltvsJ1}(b)], despite the concurrence being tiny and at first glance the concurrence still looks like in Fig. \ref{fig:DltvsJ1}(a), but the threshold temperature is highly different  compared to that without magnetic field, such as  there
is no threshold temperature for $J_{1}/J<-1$ and $\Delta\gtrsim-0.3$.

\begin{figure}
\includegraphics[scale=0.3]{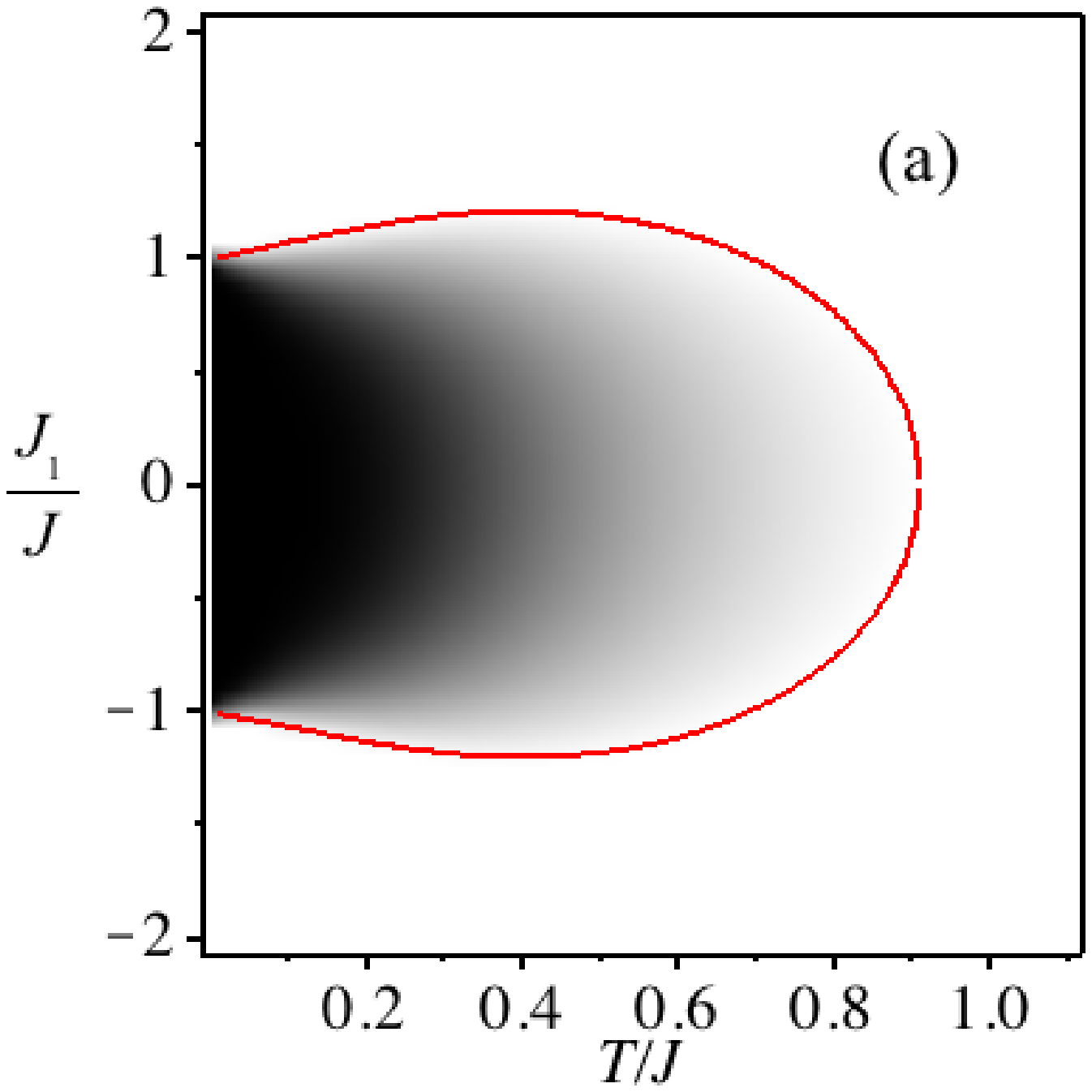}\includegraphics[scale=0.3]{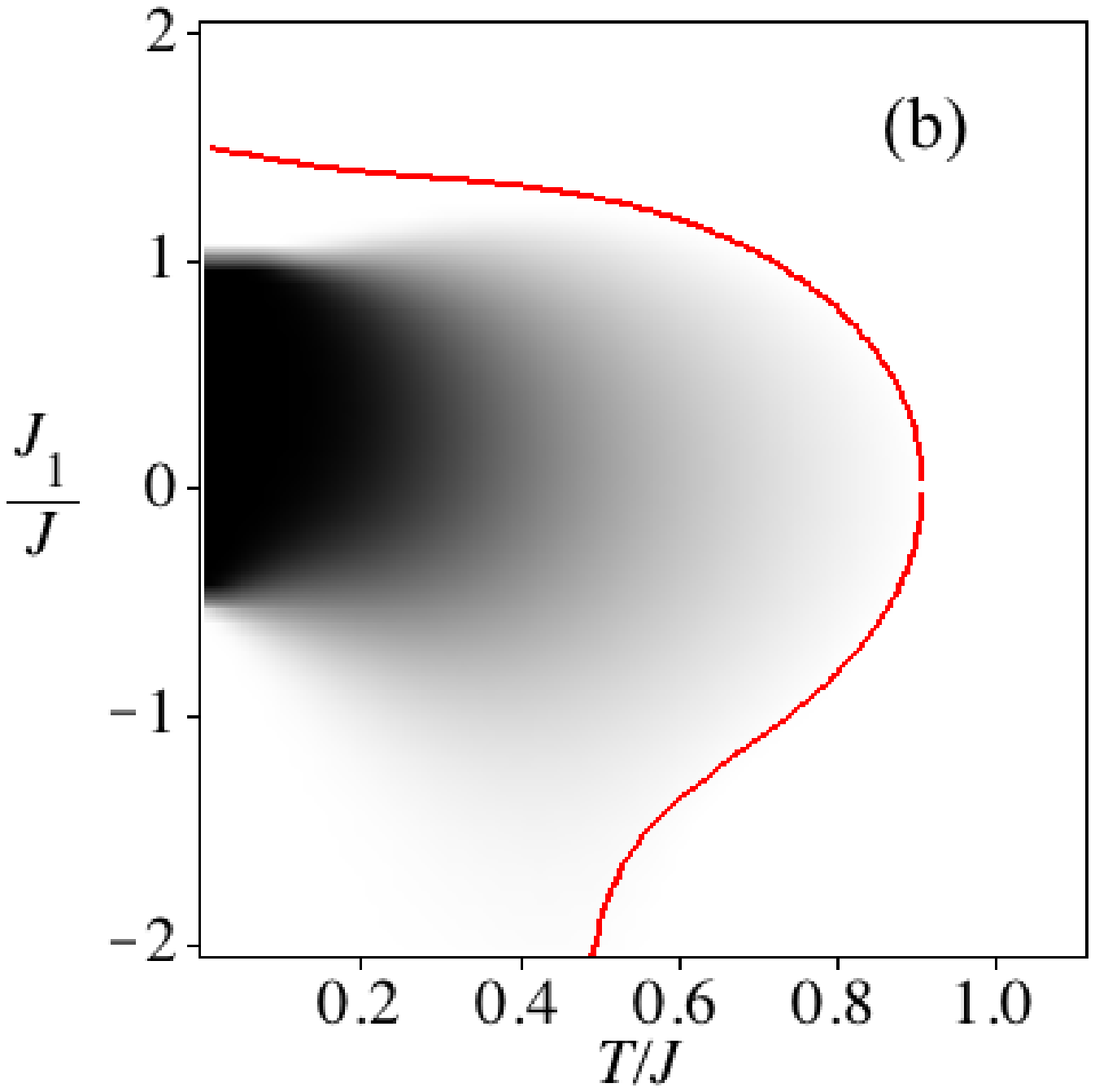}\caption{\label{fig:TvsJ1}(Color online) Density plot concurrence $\mathcal{C}$ as function of $T/J$
versus $J_{1}/J$. In (a) we display $\mathcal{C}$ for
a $h/J=0$ and in (b) we display $\mathcal{C}$ for $h/J=1$.}
\end{figure}

The density plot of concurrence $\mathcal{C}$ as a function of $T/J$ versus $J_{1}/J$
is depicted in Fig. \ref{fig:TvsJ1}, for a fixed value of $\Delta=1.0$.
We show again that the maximally entangled region vanishes for higher
temperature $T/J\approx0.3$. In Fig. \ref{fig:TvsJ1}(a) the density plot of concurrence
$\mathcal{C}$ is displayed for $h/J=0$, and the maximally entangled
region is limited by $|J_{1}/J|\lesssim1$, following the same kind of density plot as in Fig. \ref{fig:DltvsJ1}, by the solid
red line we represent the threshold temperature limiting the entangled
region. Whereas for $h/J=1$, as displayed in Fig. \ref{fig:TvsJ1}(b),
the maximally entangled region shrunk to $-0.5\lesssim J_{1}/J\lesssim1$,
despite the entangled region spread out with weakly entangled region
for $J_{1}/J<0$ and $T_{d}/J\lesssim0.5$, consequently the symmetry
in $J_{1}/J\leftrightarrow-J_{1}/J$ is broken, this is better illustrated
as a function of the threshold temperature curve described by the red solid
line.

\section{Conclusions}
In general most of  the quantum entanglement studied of an infinite chain is a cumbersome issue, motivated by this fact we investigate the quantum entanglement of the Ising-XXZ diamond chain.
Even when Ising coupling does not contribute directly in quantum entanglement,
we discuss the Heisenberg two-qubit entanglement effect on Ising-Heisenberg
diamond chain structure. First we obtain the Heisenberg dimer operators
immersed in a diamond chain with Ising coupling. Using this result we
are able to obtain the average of the two-qubit operator. Thereafter, the concurrence is obtained straightforwardly
in terms of the reduced density matrix operator elements. Using the concurrence, we study
the entanglement of the Ising-XXZ diamond chain as a function of Hamiltonian
parameters, such as  temperature as well as external magnetic field.

Usually the entangled region vanishes when the temperature increases; in some
cases [see for instance Figs. \ref{fig:DltvsJ1}(b) and \ref{fig:TvsJ1}(b)]
the entanglement vanishes asymptotically  when the external magnetic field is switched
on. The entangled region is limited by the so-called threshold temperature, but for
some other parameters [like in Figs. \ref{fig:DltvsJ1}(b) and \ref{fig:TvsJ1}(b)]
the threshold temperature only occurs in the asymptotic limit.  From our result we conclude that there is no double peak in the concurrence as a function of temperature, such as that obtained by the use of the cluster approach\cite{ananikian} [see Fig. \ref{fig:hvsC}(b)]; the disappearance of the tiny peak can be understood as a dissipation due to the qubits are inmerse in diamond chain structure. However this small peak is irrelevant, since, the degree of entanglement is rather small: $\mathcal{C}=0.01$.

It would be interesting also in the future to consideration, the case of tripartite entanglement for Heisenberg
coupling of an Ising-Heisenberg chain instead of bipartite Heisenberg coupling. Somewhat similar to that considered by
 Tsomokos et al. \cite{Tsomokos} where the tripartite coupling was studied at the zero temperature.

\section*{Acknowledgment}

O. R. and S. M. de Souza thank CNPq and Fapemig for partial financial
support. This work was been supported by the French-Armenian Grant
No. CNRS IE-017 (N.A.) and by the Brazilian FAPEMIG Grant No. CEX
BPV 00028-11 (N. A.).

\end{document}